\documentclass[trackchanges,twocolumn,twocolappendix]{aastex701}

\begin{document}

\title{Simulating Star Formation and Star Cluster Assembly in the Aquila Rift Using Archival Observations}

\author[0000-0003-3328-329X]{Jeremy Karam}
\affiliation{Department of Astronomy, Graduate School of Science, The University of Tokyo, 7-3-1
Hongo, Bunkyo-ku, Tokyo 113-0033, Japan}
\email[show]{jeremyk@g.ecc.u-tokyo.ac.jp}

\author[0000-0002-6465-2978]{Michiko S. Fujii}
\affiliation{Department of Astronomy, Graduate School of Science, The University of Tokyo, 7-3-1
Hongo, Bunkyo-ku, Tokyo 113-0033, Japan}
\email{fujii@astron.s.u-tokyo.ac.jp}

\author[0000-0001-7594-8128]{Rachel Friesen}
\affiliation{David A. Dunlap Department of Astronomy \& Astrophysics, University of Toronto, 50 St. George St., Toronto, ON, M5S 3H4, Canada}
\email{friesen@astron.utoronto.ca}

\author[0000-0003-3551-5090]{Alison Sills}
\affiliation{Department of Physics and Astronomy, McMaster University, 1280 Main Street West, Hamilton, ON, L8S 4M1, Canada}
\email{asills@mcmaster.ca}

\begin{abstract}

We simulate star formation and star cluster assembly inside a molecular cloud with parameters we derive directly from observations of the Aquila Rift. We model the evolution of stars and gas together while resolving close encounters between stars, the formation of new stars, and stellar feedback to follow cluster formation up to the expulsion of the surrounding gas. We find that star formation takes place in clumps spaced unevenly along Serpens South and that these clumps accrete surrounding gas to grow and form new stars. Gas flows along the filament promote the merger of these clumps into a star cluster inside the Serpens South filament. The imprints of these mergers are seen in the dynamics of the Serpens South cluster in the form of velocity space anisotropies, cluster rotation, and cluster expansion. Before gas is removed from the simulation, the Serpens South cluster merges with the nearby cluster W40 non-monolithically resulting in a fractal cluster at the end of the simulation. The dynamics inherited from the mergers throughout the simulation are still seen in the final bound stellar system after the gas has been removed. We compare these results with recent observations of Milky Way clusters to comment on their formation histories. We also study how our results change when lowering the mass resolution of our simulation and removing observations of dense gas tracers from our initial condition setup. Each of the three simulations result in different final cluster configurations pointing towards the importance of gas in cluster assembly.

\end{abstract}



\section{Introduction} \label{sec:intro}

Star formation takes place embedded inside regions rich in molecular hydrogen (H$_2$) across galaxies (\citealt{ladalada}). Although this has been understood for some time, it is still not fully clear how exactly these clouds evolve to form the star clusters we see today. Observations of molecular clouds show that they are not homogeneous in their phase space distribution but they do have similarities that connect them. An important push towards this conclusion came from high resolution observations of the molecular clouds along the Gould Belt in the Milky Way performed using the \textit{Herschel} space telescope (\citealt{HGBS}). Analysis of these clouds show that they contain networks of high density filaments (\citealt{arzoumanian_2019}) and that the formation of young stellar objects (YSOs, the precursors of stars) happens preferentially along these filaments (\citealt{konyves_2015}). To get a full picture of the star formation process, however, it is necessary to complement the \textit{Herschel} observations with spectral observations of the gas in these clouds to understand their dynamics. An early example of this analysis for the Serpens South Filament in the Aquila rift was carried out by \citet{kirk} where the authors find evidence for gas flow along the filament to a central proto-star cluster. This was further shown in higher resolution observations of the region from \citet{friesen_2024}. This type of gas flow has been observed in many filament systems (e.g. \citealt{chen_michael_2020}, \citealt{wang_2020}, \citealt{rawat}, \citealt{zhang_w49}) with some studies also finding evidence that gas flows along filaments can lead to accretion of gas by protostellar clumps (e.g. \citealt{yuan_2018}, \citealt{olguin_2021}, \citealt{beuther_2025}). A full phase space analysis of the nearby Orion integral-shaped filament finds evidence that the kinematics and morphology of the dense gas can influence greatly the positions and motions of the young protostars in the region (\citealt{stutz_2016}). 

The influence of gas morphology and dynamics on cluster assembly has been seen in simulations that follow the formation of star clusters beginning from molecular gas clouds. As stars continue to form inside molecular clouds, they begin to congregate into small subclusters that can continue to accrete surrounding gas (to form new stars), and eventually merge with one another leading to cluster growth inside the cloud hierarchically (e.g. \citealt{vazquez}, \citealt{Howard2018}, \citealt{gonzalez_2020}, \citealt{dobbs_22}). Hierarchical star cluster build up may also be the cause of the observed fractality of star clusters (e.g. \citealt{hetem_2015}, \citealt{pang_2022}, \citealt{coenda_2025}) as star cluster mergers do not always lead to monolithic clusters (\citealt{karam}, \citealt{Karam2024}). Because gas remains the dominant component of the overall potential at these early times, its kinematics can promote subcluster mergers (\citealt{Karam2024}) and impart unique dynamics onto the clusters that eventually emerge from these molecular clouds (e.g. \citealt{armstrong_tan}, \citealt{karam_25}). Depending on the gas density and motions of a cluster, movement of a star cluster through the ambient molecular gas medium can even result in changes in the bound mass components of the cluster (\citealt{karam_23}). This becomes a very important component of star cluster assembly when you consider the varying cloud phase space morphologies within which clusters are expected to form that are revealed by observational surveys (see for example the range of cloud properties in the \citealt{bazzi_2025} sample of extragalactic clouds).

Along with the \textit{Herschel} survey of the Gould Belt clouds, high resolution observations of gas morphology and kinematics from the Atacama Large Millimeter Array (\textit{ALMA}) have showcased the forms which star forming clouds can take in high resolution detail. The ALMA-IMF survey, for example, (see \citealt{motte_2022}) observed 15 potentially star cluster forming clouds with masses in the range of 10$^3$ to 10$^4$ M$_\odot$ at different evolutionary stages towards cluster formation (\citealt{galvan_madrid_2024}). While the turbulence spectrum (and in turn, the global kinematic structure) may be similar among these clouds (\citealt{koley_2025}), the mass budget held by protostellar cores within clouds classified in the same evolutionary stage can vary (\citealt{motte_2022}). This further showcases the dependence of star formation (and star cluster build up) on the resolved dynamics of the gas in a given molecular cloud. It is therefore crucial for simulations to move towards initial conditions inherited directly from observations to ensure they are modeling gas phase space morphology accurately.

This process has been done in the past using observations of the Orion integral filament to set up simulation initial conditions in \citet{sills_2023}. The authors used data of the Orion filament from the Green Bank Ammonia Survey (\citealt{friesen_2017}) which measured the emission of NH$_3$ in the region. By only considering NH$_3$, the authors were only able to capture the dynamics of the high density gas in the region (\citealt{shirley_2015}). As well, by not including stellar feedback effects, they were unable to follow the formation of the star cluster up until it expels its surrounding gas. To build upon this previous work, we add these mechanisms in our simulations with the inclusion of multiple gas tracers and a stellar feedback prescription and simulate the evolution of a different region.

The Aquila rift is a star forming molecular cloud with a young cluster forming in the dense, Serpens South filament (\citealt{kirk}) and an already formed young cluster W40 that is thought to be influencing the evolution of Serpens South through ongoing stellar feedback (\citealt{shimoikura_2019}). Serpens South is currently forming a protocluster (as can be seen by the collection of YSOs the hub of the filament) and is surrounded by a diffuse gas distribution that may be accreting onto the central filament.

In this paper, we expand on the method introduced in \citet{sills_2023} to convert observations of star forming clouds into simulation initial conditions using multiple gas tracers. We apply this method to the Serpens South filament and the neighboring young cluster W40 to study the star formation and star cluster build up process in the region in detail. This paper is organized as follows: in Section \ref{sec:methods} we discuss the conversion process and the observations used, in Section \ref{sec:results} we discuss our results related to star formation, star cluster build up, and the dynamics of the resultant cluster within the Serpens South filament, in Section \ref{sec:compare_obs} we compare our results to observations, and in Section \ref{sec:summary} we summarize our key results. In the appendix Section \ref{sec:var_ics} we discuss the dependence of our results on the mass resolution of the simulation, and datasets used to initialize simulations.

\begin{figure*}
    \centering
    \includegraphics[width=1\linewidth]{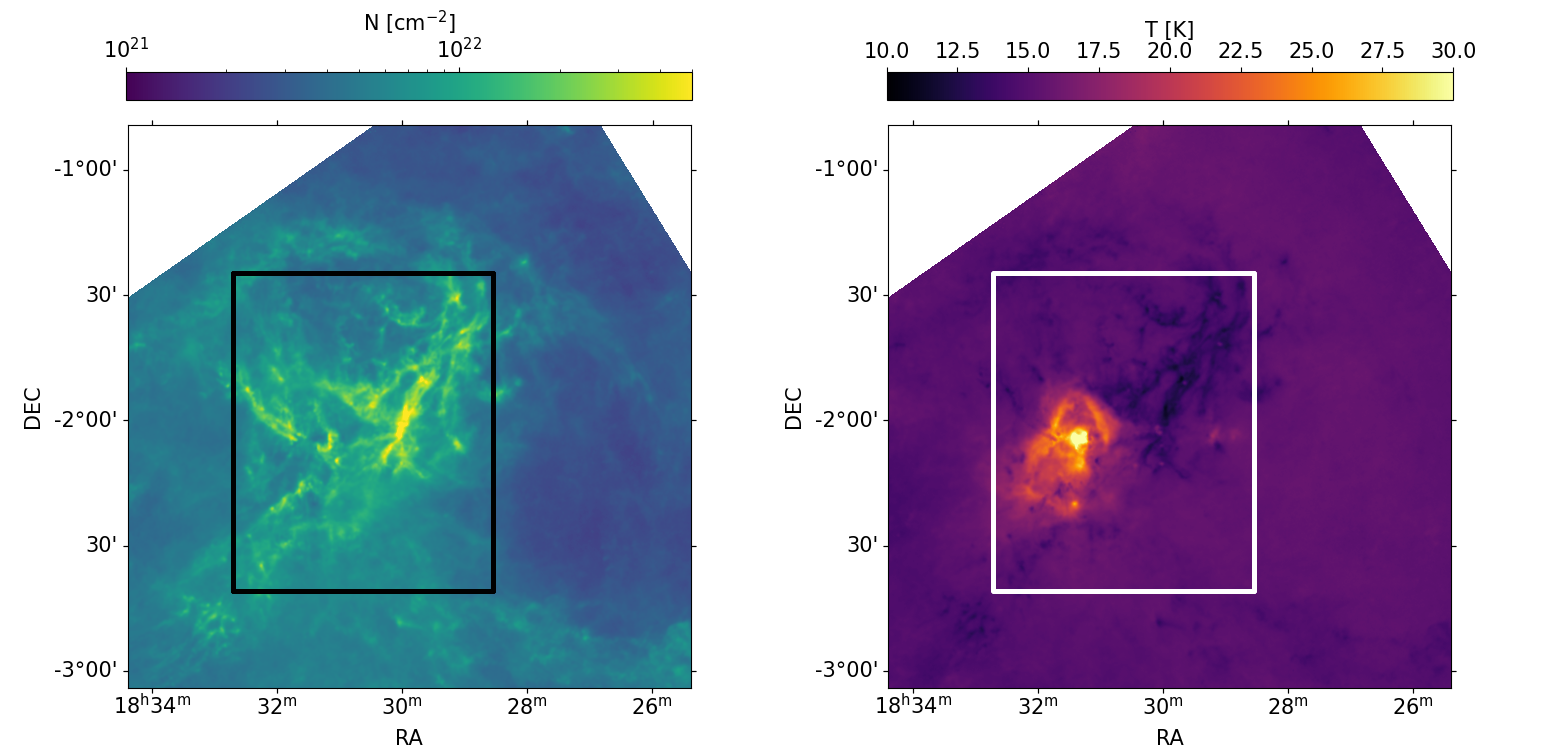}
    \caption{Observations of the Aquila star forming region from the Herschel Gould Belt Survey. Black and white boxes show region with corresponding ammonia observations from the Green Bank Ammonia Survey (see Section \ref{sec:NH3}) and the regions simulated in this work. Left: number density of H$_2$. Right: Temperature map of H$_2$. }
    \label{fig:aquila_herschel}
\end{figure*}

\section{Methods} 
\label{sec:methods}

\subsection{Numerical Methods}
To perform our numerical simulations, we use the ASURA+BRIDGE code (\citealt{fujii_sirius_2}, \citealt{fujii_sirius_3}). ASURA+BRIDGE is a star-by-star code that evolves gas using the smoothed particle hydrodynamics (SPH) code \texttt{ASURA} (\citealt{saitoh_2008}, \citealt{saitoh_2009}) and stars using the N-body dynamics code \texttt{PETAR} (\citealt{Wang2020a}) which allows us to follow the evolution of dynamically formed binary and multiple systems (\citealt{Wang2020b}). Evolving both gas and stars together requires a connecting scheme such that the two components respond to changes in each other. This is done through \texttt{BRIDGE} (\citealt{bridge}). The bridge timestep determines how often the N-body and hydrodynamics codes are connected and we choose it to be 200\,yr as determined through test simulations in \citet{fujii_sirius_3}. The softening length for the gas is 500\,AU and the softening length for the stars is 0pc. This choice allows us to resolve the formation of dynamical binaries and higher order multiple systems thus accurately modeling the dynamics of the cluster. We refer the reader to \citet{Wang2020a} for more details on the method.


The star formation prescription included in ASURA+BRIDGE follows the formation of individual stars throughout the simulation. A description of the star formation prescription can be found in \citet{fujii_sirius_1} and we summarize it here. Once gas is above a chosen density threshold, below a chosen temperature, and is converging, it is eligible to form stars. We choose density and temperature thresholds to be 10$^5$cm$^{-3}$, and 30K respectively. This density threshold is similar to the value of observed prestellar cores (\citealt{enoch_2008}) and the temperature threshold is in the range of observed star forming molecular clumps (\citealt{guzman_2015}). A stellar mass is then sampled from a Kroupa (\citealt{kroupa2001}) IMF stochastically and mass is assembled from the eligible gas mass to equal the chosen stellar mass. The star formation efficiency is set to 0.02 to match observational constraints (\citealt{krumholz_19}). In total, the star formation prescription depends on the star formation efficiency, and the maximum search radius within which gas is sampled to form stars. For a complete survey on the effects of varying these parameters, we point the reader to \citet{fujii_sirius_1}.

Once the necessary amount of gas is collected, it is replaced by an N-body stellar particle whose position and velocity are inherited from the gas from which it formed. Star particles do not accrete throughout the simulation and mass loss is calculated using \citet{hurley_2000}.

Feedback in our simulations is modeled using Str{\"o}mgren radii (\citealt{stromgren}) around massive stars. Local density and ionizing photon rates are used to calculate these radii using \citet{lanz_hubeny}. Gas within the Str{\"o}mgren radius of a given star is injected with thermal energy and momentum. A description of the feedback prescription can be found in \citet{fujii_sirius_3}.

To set up the simulations presented in this paper, we take archival observations of the Aquila star forming region from multiple telescopes and convert the maps into distributions of SPH and N-body particles. Each SPH particle in our simulations requires a mass ($m_{\mathrm{SPH}}$), 3D position (x, y, and z), 3D velocity (v$_\mathrm{x}$, v$_{\mathrm{y}}$, and v$_{\mathrm{x}}$), and internal energy ($u$) while each N-body particle requires a mass, 3D position, and 3D velocity.

\subsection{Observational Data}

In this section, we discuss how we derive the necessary parameters for our initial conditions from archival observations of the Aquila star forming region. The region we simulated contains the W40 star cluster along with the young Serpens South protocluster, and encloses all datasets we use. Throughout this process, we assume an inclination angle of 0$^{\circ}$ (entirely face-on).

\begin{figure*}
    \centering
    \includegraphics[width=1\linewidth]{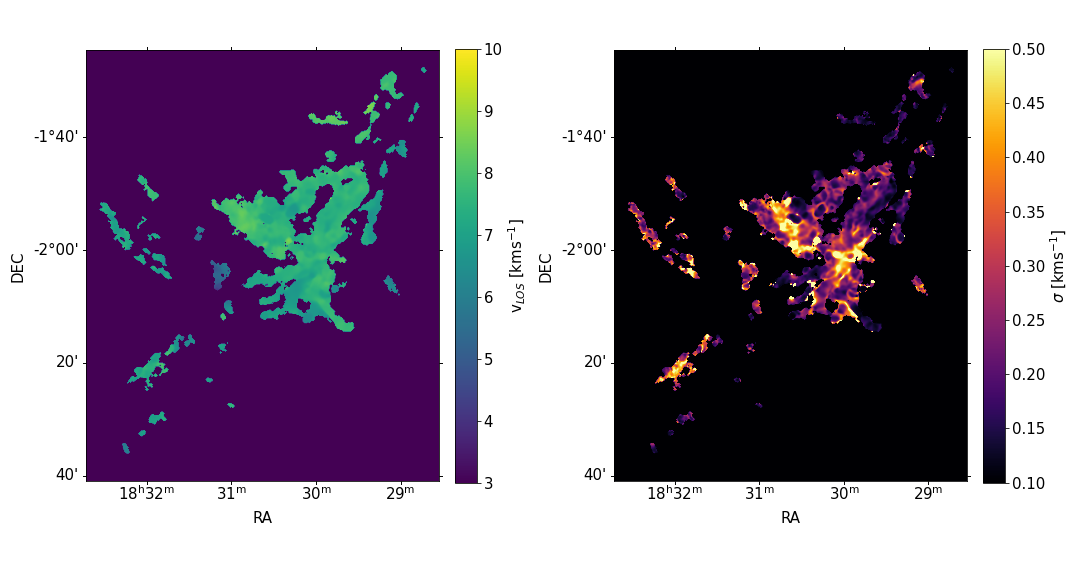}
    \caption{Observations of the Aquila star forming region from the Green Bank Ammonia Survey. Left: line of sight velocities of NH$_3$. Right: Velocity dispersion of NH$_3$.}
    \label{fig:greenbank_aquila}
\end{figure*}

\subsubsection{Herschel: Mass, Position, and Internal Energy}
\label{sec:herschel}
We begin with the column density ($N$$_{\mathrm{H}_{\mathrm{2}}}$) and dust temperature ($T$) maps of H$_\mathrm{2}$ provided by the Herschel Gould Belt Survey (HGBS)\footnote{\url{http://www.herschel.fr/cea/gouldbelt/en/Phocea/Vie_des_labos/Ast/ast_visu.php?id_ast=66}} (\citealt{HGBS}). They can both be seen in Figure \ref{fig:aquila_herschel}. These maps are derived from the Herschel observations of the region in \citet{konyves_2015} and we summarize the process here.

The maps are derived by combining the images of surface brightness at 160$\mu$m, 250$\mu$m, 350$\mu$m, and 500$\mu$m and fitting a modified black body to each pixel in each image. For this fit, the authors used a power law approximation to the dust opacity using $\kappa_\lambda = 0.1 \times (\lambda/300\,\mu \mathrm{m})^{-\beta}$\,cm$^2$g$^{-1}$ with a fixed dust emissivity index $\beta = 2$ (\citealt{hillebrand_1983}). In this fitting procedure, the dust surface density ($\Sigma$) and dust temperature are calculated. The H$_\mathrm{2}$ column density was then derived using $\Sigma = \mu_{\mathrm{H_2}}m_\mathrm{H}$$N$$_{\mathrm{H}_{\mathrm{2}}}$ where $\mu_{\mathrm{H_2}}=2.8$ is the mean molecular weight per hydrogen molecule, and $m_{\mathrm{H}}$ is the mass of one hydrogen molecule. 

From the column density map shown in the left-hand panel of Figure \ref{fig:aquila_herschel}, we derive the mass of H$_\mathrm{2}$ in each pixel using $M$$_{\mathrm{H_2},pix}$=$A$$_{pix}$$\mu_{\mathrm{H_2}}m_\mathrm{H}$$N$$_{\mathrm{H_2},pix}$ where A$_{pix}$ is the area of the \textit{Herschel} pixel, and $N_{\mathrm{H_2},pix}$ is the column density of H$_{\mathrm{2}}$ in each \textit{Herschel} pixel. To find the area of each pixel, we use a distance of $d$$=436$pc to the Aquila region (\citealt{ortiz_leon_2018}).

Once we have the mass of H$_{\mathrm{2}}$ in each pixel, we place $n_{\mathrm{H_2}} = M_{\mathrm{H_2},pix}/m_{\mathrm{SPH}}$ particles, where $m_{\mathrm{SPH}} = 0.005$M$_{\odot}$, randomly in each pixel. We discuss this choice in Section \ref{sec:IC_iters}. Because we do not have information regarding the distribution of gas along the line of sight, we sample the z position of each particle that belongs to a filament randomly between $-0.1$\,pc and $+0.1$\,pc as this is the average filament width of the filaments present in the Aquila region (\citealt{arzoumanian_2011}, \citealt{arzoumanian_2019}). We determine whether or not gas belongs to a filament using a surface density cut of 6$\times$10$^{21}$\,cm$^{-2}$ where gas above this density is considered filamentary. This threshold corresponds to the density at the edge of the Aquila filaments as determined in \citet{arzoumanian_2019}. Particles that are not part of a filament are sampled randomly between a z-distance of $-18$\,pc and $+18$\,pc. We determine this value using the \citet{edenhofer_map} dust map to determine the column density of dust along the line of sight pointed in the direction of the Aquila rift ($l \approx 28.8^{\circ}$ and $b \approx 3.5^{\circ}$).

We convert the temperature maps from the HGBS into internal energy maps using

\begin{equation}
    u_{pixel} = \frac{3k_BT_{pixel}}{2\mu_{H_2}}
\end{equation}
where u$_{pixel}$ is the internal energy in each pixel, k$_B$ is the Boltzmann constant, and T$_{pixel}$ is the temperature at each pixel. Each SPH particle is given the internal energy corresponding to the pixel within which it is placed.

\subsubsection{Green Bank: Dense Gas Kinematics}
\label{sec:NH3}
The Green Bank Ammonia Survey (\citealt{friesen_2017}) performed spectroscopy of the NH$_3$ gas present in the Aquila region providing NH$_3$ line-of-sight velocities and velocity dispersions as seen in Figure \ref{fig:greenbank_aquila}. Observations of NH$_3$ gas correspond to the dense components of a given molecular cloud.

To incorporate these maps in our simulation initial conditions we first calculate the amount of NH$_3$ mass in a given pixel using the column density maps of NH$_3$ and
\begin{equation}
\label{eq:mass}
    M_{GB,pix} = A_{GB,pix}\frac{N(\mathrm{NH_3})_{GB,pix}}{X(\mathrm{NH_3})}\mu_\mathrm{H_2} m_{\mathrm{H}}
\end{equation}
where $M_{GB,pix}$ is the mass traced by NH$_3$ in each \textit{Green Bank} pixel, $A_{GB,pix}$ is the area of each \textit{Green Bank} pixel, $N(\mathrm{NH_3})_{GB,pix}$ is the column density of NH$_3$ in each \textit{Green Bank} pixel, and $X(\mathrm{NH_3})$ = 1.5$\times$10$^{-8}$ is a conversion factor between H$_2$ column density and NH$_3$ column density in accordance with \citet{friesen_2024}. We then use this to derive the number of SPH particles traced by NH$_3$ per pixel using $n_{GB,pix}$ = $M_{GB,pix}/m_\mathrm{SPH}$.

One \textit{Green Bank} pixel is $\approx$ 3 times larger than a given \textit{Herschel} pixel implying that within one \textit{Green Bank} pixel is $\approx$ 9 \textit{Herschel} pixels. To determine which SPH particles are considered NH$_3$, we randomly select $n_{GB,pix}$ SPH particles within a given \textit{Green Bank} pixel. We then use the line-of-sight velocity, and velocity dispersion from the corresponding \textit{Green Bank} pixel to infer the velocities of the SPH particles traced by the NH$_3$ observations. To do this, we sample the z-component velocity of each particle from a Gaussian centred on the line-of-sight velocity of the given pixel with a width given by the velocity dispersion of the same pixel. We then assume cylindrical symmetry throughout the simulation and assign each particle an x and y velocity equal to its z-component velocity as is done in \citet{sills_2023}. While other options are available to assign velocities along x and y, \citet{sills_2023} showed that they have little overall affect on the simulation outcome.

\subsubsection{Nobeyama Radio Observatory: Kinematics}
\label{sec:nobeyama}
To provide our remaining SPH particles with initial velocities, we turn to the spectral observations of Aquila from the Nobeyama Radio Observatory (NRO) (\citealt{shimoikura_2020}). We take the position-position-velocity (PPV) cubes of $^{13}$CO from the publicly available online database of the NRO star formation project archive\footnote{\url{https://jvo.nao.ac.jp/portal/nobeyama/sfp.do}}. We use the \texttt{specutils} package in \texttt{python} to fit single-component Gaussian spectra to each pixel in the PPV cube and take the mean of the Gaussian as the line-of-sight velocity of the gas in each pixel, and the standard deviation of the Gaussian as the velocity dispersion of each pixel. The fit returns a covariance matrix which we use to determine the errors in the line-of-sight velocities and velocity dispersions in each pixel. In the Aquila rift, there is evidence for double component line-of-sight velocities in the $^{13}$CO maps, but the mass of gas with this signature is very small compared to the total mass of the system in our simulation (\citealt{shimoikura_2020}).

Particles that are not considered dense gas (as calculated using the Green Bank NH$_3$ maps described in Section \ref{sec:NH3}) have their velocities inferred from the NRO maps. Similar to velocities assigned for dense gas, we randomly sample the z-component velocity from a Gaussian centred on the line-of-sight velocity of the given pixel as well as the velocity dispersion of the same pixel. The x and y components of the velocity are assigned the same value as the z-component velocity to maintain cylindrical symmetry. There are pixels of the map where the signal is dominated by noise. We give gas particles belonging to these pixels a line-of-sight velocity value equal to the average line-of-sight velocity of the entire map and a velocity dispersion equal to the average velocity dispersion of the map. As there are only few such pixels, they have little effect on our results.

As seen by the high temperature gas in the right panel of Figure \ref{fig:aquila_herschel}, there is an H\textsc{ii} region to the left of the Serpens South filament that is heating up the surrounding molecular gas. This region is produced by stars found in a young star cluster known as W40 (e.g. \citealt{rodney_2008}, \citealt{mallick_2013}) that we will discuss further in Section \ref{sec:stellar}. As modeling this H\textsc{ii} region is not the goal of this study, we elect a simple approach based on the analysis of the region from \citet{shimoikura_2019} where the authors determined that the H\textsc{ii} region has a pinched hourglass shape with an outer diameter of $\approx 2.5$pc. The velocity of the gas inside the region was determined to be $\approx 3$\,kms$^{-1}$. We therefore set up two spheres above and below the centre of W40 each with a diameter of 2.5\,pc and give all the gas within these spheres a velocity of 3\,kms$^{-1}$ directed away from the cluster centre. 

Our method inherently includes the turbulence spectrum of the Aquila region up to the resolution limit of the observations we use by mapping the velocities directly onto the simulation conditions. Incorporating higher resolution observations will require a more in depth modeling of protostar formation and evolution including disk physics and accretion and will be left to future work.

\begin{center}
\begin{table}[]
\hspace{-1.5cm}
    \centering
\begin{tabular}{c|c|c}
    Model Name & Observational Datasets Used  & m$_{SPH}$[M$_\odot$]\\
    \hline 
     \texttt{fiducial} & \textit{Herschel}, \textit{Green Bank} &  0.005\\
     \texttt{low\_res} & \textit{Herschel}, \textit{Green Bank} &  0.01 \\
     \texttt{no\_NH3} & \textit{Herschel} &  0.005
     
\end{tabular}
    \caption{Simulation parameters. Column 1: simulation name, column 2: gas observations used for particle velocities, column 3: SPH particle mass.}
    \label{tab:sims}
\end{table}
\end{center}

\begin{figure}
    \centering
    \includegraphics[scale=0.32]{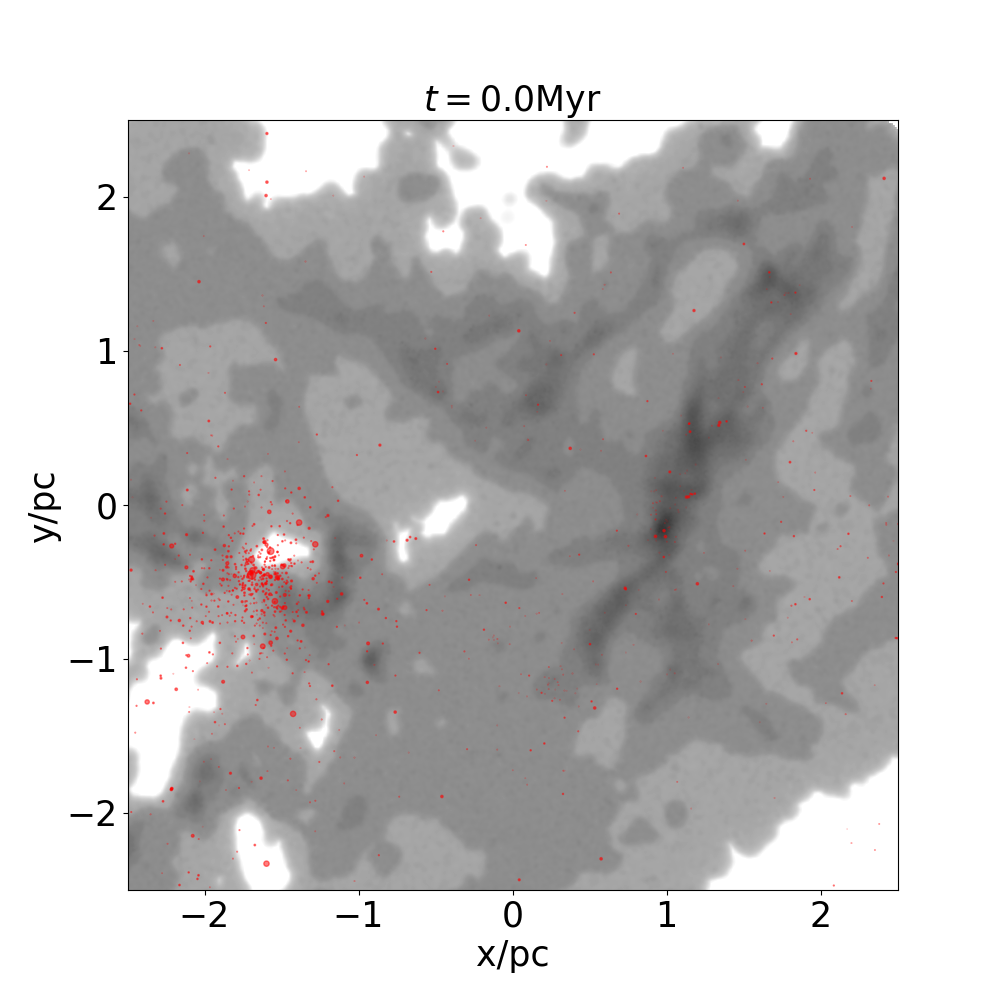}
    \caption{Initial condition iteration used in the \texttt{fiducial} simulation in this work. Grayscale shows the gas density, and red points show star particles with their size proportional to their mass in the mass range of 0.01M$_\odot$ to 18.7M$_\odot$.}
    \label{fig:ics}
\end{figure}

\subsubsection{Pre-Stellar and Stellar Components}
\label{sec:stellar}
The Aquila region hosts a rich population of class I and class II young stellar objects (YSOs) along with more evolved stars, one of which is driving the HII region in W40 located to the east of the Serpens South filament (\citealt{shuping_2012}). We model both YSOs and stars as N-body particles in our simulation using the method outlined in this section.

We use the catalogue of class I and class II YSOs presented in \citet{sun_2022} to populate our simulation with initial YSO particles. \citet{sun_2022} collected near-infrared observations from the Canada-France-Hawaii Telescope (CFHT), the Two Micron All Sky Survey (2MASS) (\citealt{skrutskie_2006}), and the UKIRT Infrared Deep Sky Survey (UKIDSS) (\citealt{lucas_2008}), coupled with mid-infrared observations from Spitzer to identify and classify YSOs in the Aquila region. In total, this catalogue provides the right ascension and declination of each YSO, along with magnitudes and magnitude uncertainties in the J, H, K, and Spitzer bands. \citet{sun_2022} also classify each YSO into class I or II.

The right ascension and declinations give us the x and y positions of each YSO in our simulation. Along the z-axis, we give each YSO a random position between $-$0.1pc and $+$0.1pc to match the dense gas distribution as they are more tightly linked to the dense gas (see Section \ref{sec:herschel}). Lastly, we assign a mass to each YSO by comparing their magnitudes to the pre-main sequence isochrones presented in \citet{baraffe_2015}. If the observed YSO is class I, we use the 0.5Myr isochrones for comparison, and if it is class II, we use the 5Myr isochrones for comparison. This matches age estimations of the YSOs in the Aquila region from previous studies (\citealt{plunkett_2018}, \citealt{sun_2022}). 

The W40 star cluster has been observed through the MYStIX project as outlined in \citet{feigelson_2013}. These observations show that W40 contains 426 stars. Further analysis of the distribution of these stars in \citet{kuhn_2014} show that W40 is spherical with a core radius of 0.17pc. We therefore distribute 426 N-body particles in a Plummer (\citealt{plummer1911}) sphere with a scale radius of determined from the core radius around the centre of the cluster as given in \citet{kuhn_2014}. We sample masses for these stars from a Kroupa (\citealt{kroupa2001}) IMF with a range of 0.08M$\odot$ to 10M$_\odot$ (the upper limit of 10M$_\odot$ is chosen because of the limits of the MYSTiX survey). Following \citet{shimoikura_2019}, we assume that the W40 cluster and its surrounding gas component are connected to the gas belonging to Serpens South and therefore place the cluster stars on the same plane as the gas in the simulation.

Along with the MYSTiX project, 29 stars belonging to the W40 cluster have been observed with the Gaia space telescope in \citet{cameron_2022}. We include these stars in our simulation by converting the right ascension and declination to x and y positions, the parallax to a z position, the proper motions into x and y component velocities, and the radial velocities to a z-component velocity. We then normalize the z position of these stars by placing the massive O star in the same plane as the gas and the W40 cluster as defined by MYSTiX because we know that this star is driving the observed HII region. For stars without radial velocity measurements, we assign them a z-component velocity equal to the mean line of sight velocity of the Aquila region. Of these 29 stars, 21 of them were accompanied with spectroscopy to determine a spectral type. We use the tables presented in \citet{pecaut_2013} to assign masses to these stars. For the remaining 8 stars, we assign them a mass of 0.7M$_\odot$ which is the average mass of the IMF chosen.

\subsection{Initial Condition Iterations}
\label{sec:IC_iters}

To study the drivers of star formation in this system, we run three iterations of the same simulation with each having slight differences. The first model (\texttt{fiducial}) is initiated as described throughout Section \ref{sec:methods} and is shown in Figure \ref{fig:ics}. We choose $m_\mathrm{SPH}=0.005$M$_\odot$ for our \texttt{fiducial} simulation because it sufficiently resolves the density structure of the region while not unnecessarily increasing computational costs. Our second model is the same as the \texttt{fiducial} model, but does not include kinematics of the dense gas (NH$_3$) inherited from the Green Bank telescope as described in Section \ref{sec:NH3}. In essence, this model is only simulating the $^{13}$CO dynamics. We call this model \texttt{no\_NH3}. In our third model, we mimic observational resolution constraints by running a model with higher $m$$_{\mathrm{SPH}}$ (lower mass resolution) titled \texttt{low\_res}. The models are summarized in Table \ref{tab:sims}.

\begin{figure*}
    \centering
    \includegraphics[scale=0.3]{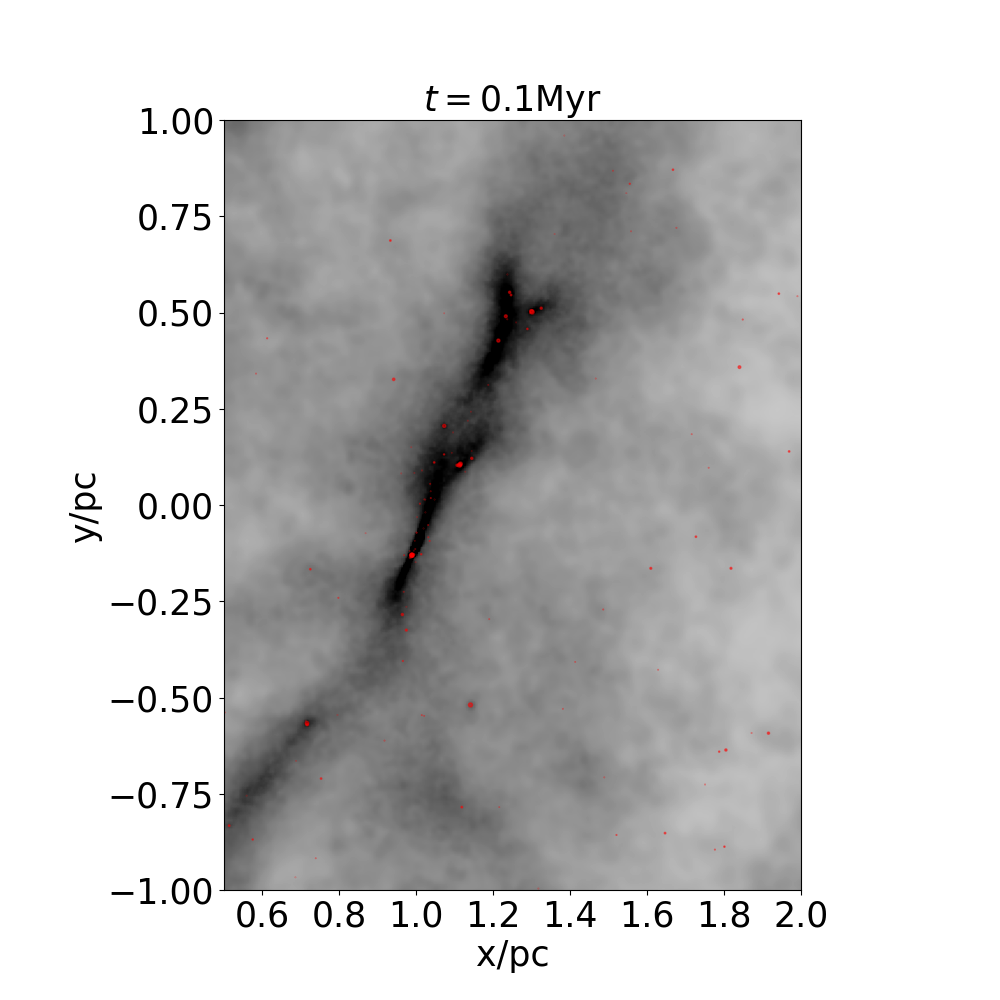}
    \includegraphics[scale=0.3]{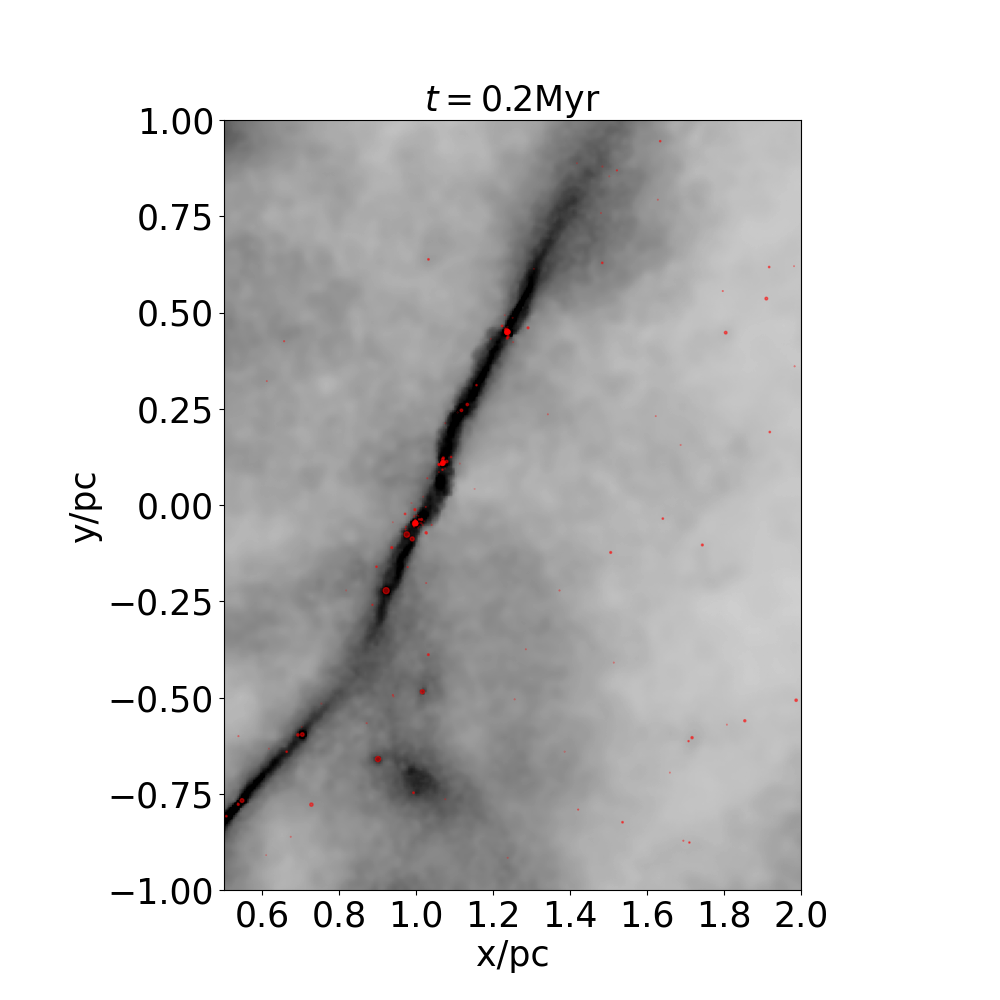}
    \includegraphics[scale=0.3]{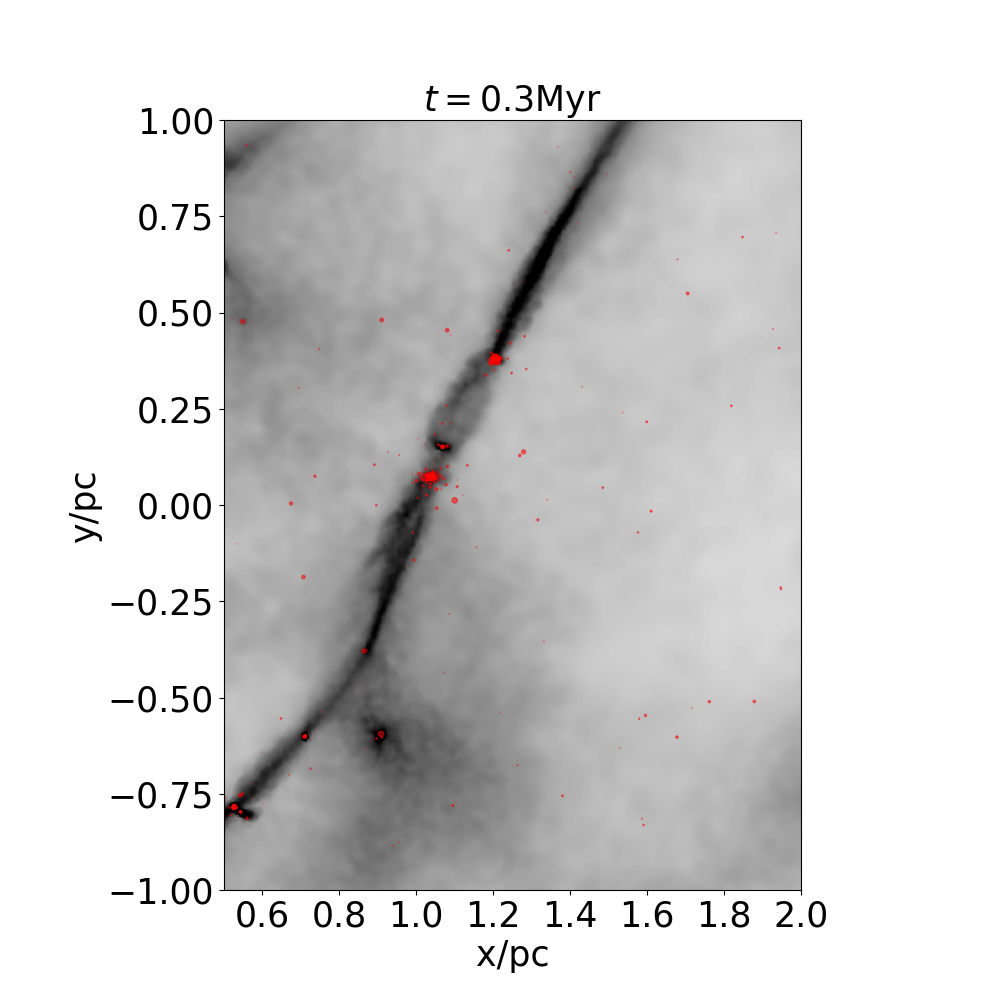}
    \includegraphics[scale=0.3]{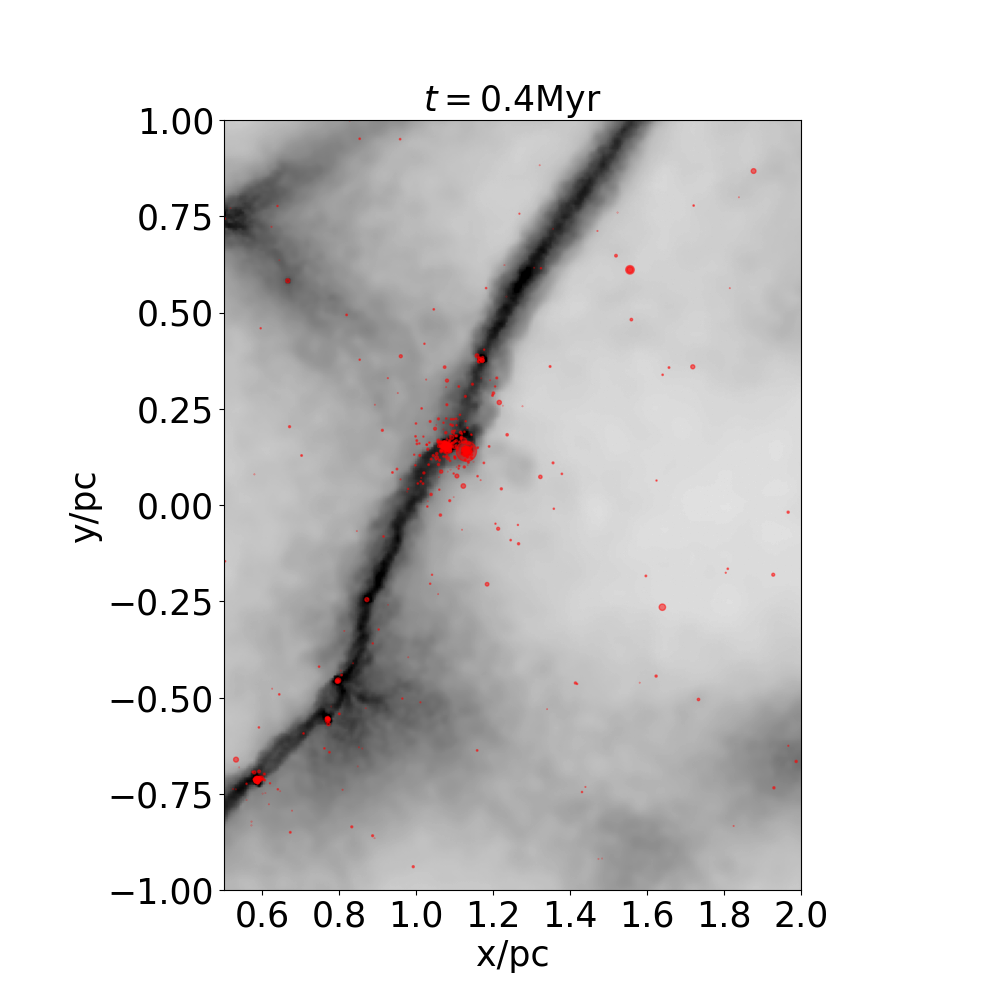}
    \caption{Snapshots showing the first 0.4Myr of evolution of the Serpens South filament from our \texttt{fiducial} simulation. The greyscale shows the gas volume density and the red points show star particles with the size proportional to the mass of the star.}
    \label{fig:serpens}
\end{figure*}

\section{Results} \label{sec:results}
In the following section, we focus our discussion on the \texttt{fiducial} model and discuss the other simulations in the Appendix. We begin our discussion with the evolution of the Serpens South filament and the star formation that occurs within it as the system evolves.

\begin{figure}
    \centering
    \includegraphics[scale=0.41]{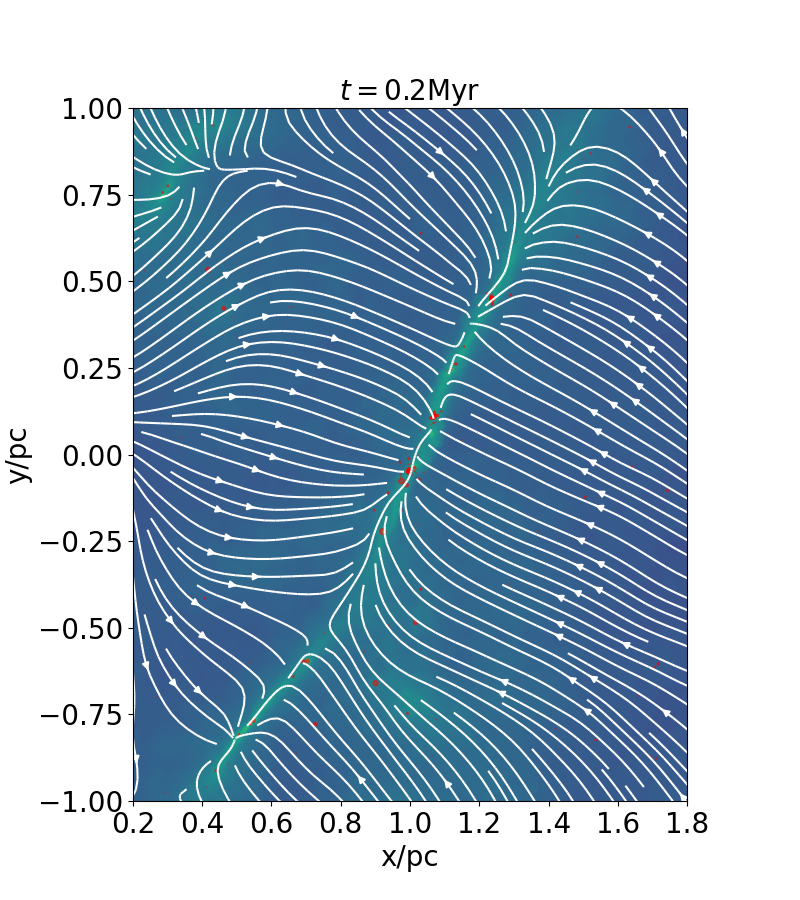}
    \caption{Same as the top right panel of Figure \ref{fig:serpens} but with streamlines overlayed showcasing the density averaged velocity of the gas around the Serpens South filament.}
    \label{fig:serpens_qual}
\end{figure}

\begin{figure}
    \centering
    \includegraphics[scale=0.3]{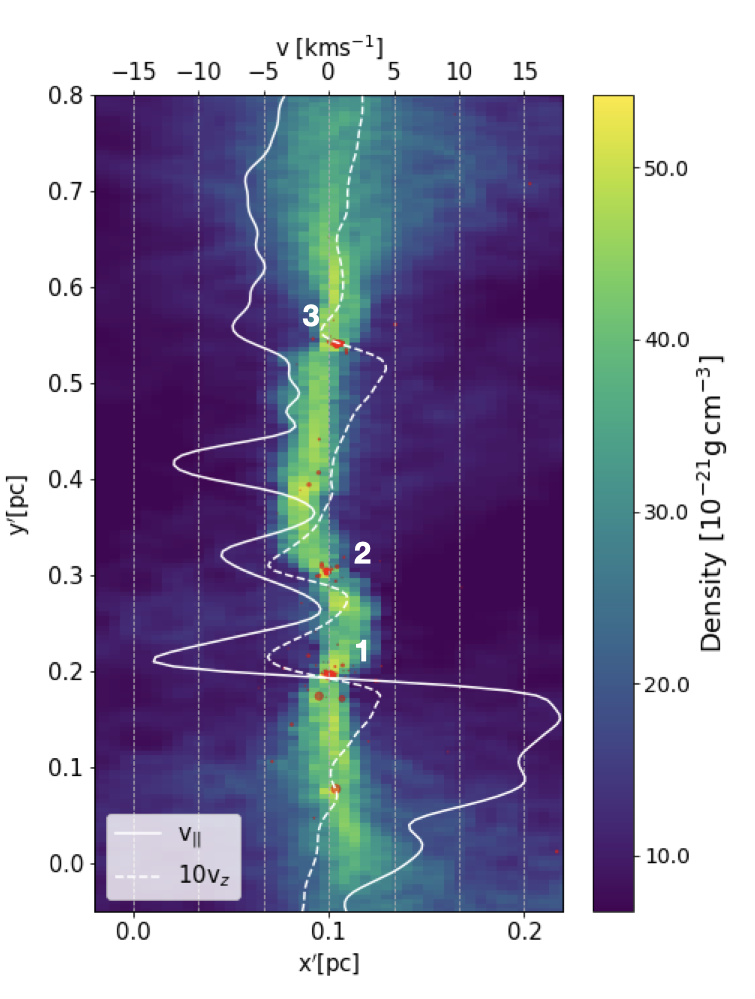}
    \caption{Dynamics of the gas around the Serpens South filament at $t = $0.2\,Myr in our \texttt{fiducial} simulation along with gas density averaged along the line of sight (in viridis) in the rotated frame such that the filament is parallel to the y$^{\prime}$-axis. The white lines show the velocity along the filament (solid) and the velocity along the line of sight multiplied by a factor of 10 to place both lines on the same scale (dashed). The red points show the locations of the stars with their size proportional to the mass of the star. Stellar clumps are labeled one through three.}
    \label{fig:serpens_dynamics}
\end{figure}

\begin{figure*}
    \centering
    \includegraphics[scale=0.4]{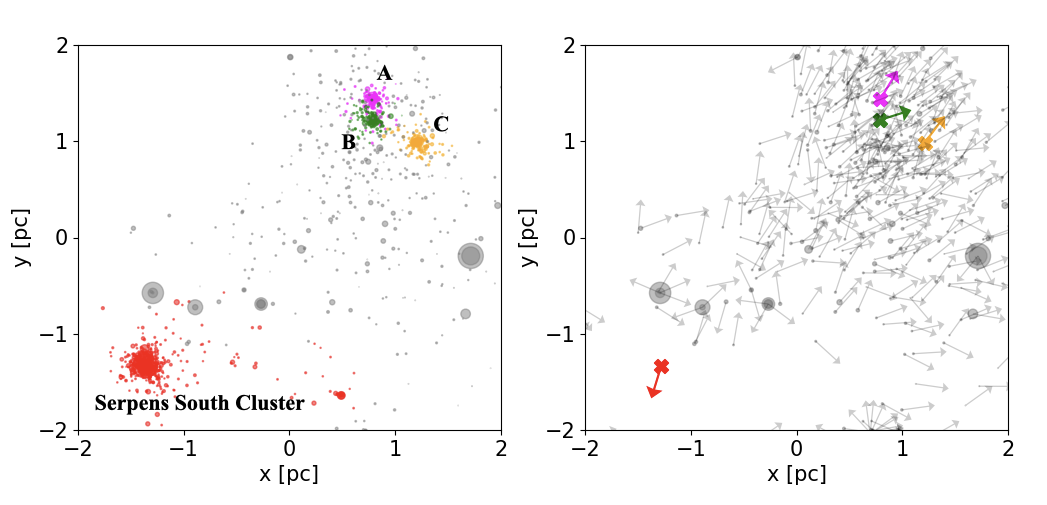}
    \caption{The bound stellar component at the time of gas removal in the \texttt{fiducial} simulation at $t=1.6$\,Myr. Left: physical distribution of bound stars. The sizes of the circles are proportional to the mass of the star. Colours show the groups identified with HDBSCAN with gray points showing unclustered stars. Right: same as left but with unit vectors showing the direction of motion of bound, unclustered stars. Clusters are replaced with their centre of mass and centre of mass velocity vectors.}
    \label{fig:final_snap}
\end{figure*}

\subsection{Star Formation Along Serpens South}
We show snapshots of the evolution of the Serpens South filament in Figure \ref{fig:serpens}. After $\approx$0.1Myr, the filament collapses along its spine and star formation accelerates. Filaments of this width have been seen in very high resolution simulations of star cluster formation presented in \citet{suin_2024}. Star formation proceeds in clumps located along the filament spine and, as seen in the middle-left panel of Figure \ref{fig:serpens}, these clumps are not equally spaced apart. This uneven spacing comes from the non-uniform density distribution present in Serpens South and inherited in our simulation initial condition. This is an example of the importance of realistic initial conditions in simulations.

As the clumps continue to form stars, they begin to accrete surrounding filamentary gas. We show this accretion qualitatively in Figure \ref{fig:serpens_qual} where we show the velocities of the gas along Serpens South as velocity streamlines overlaid on top of the gas density within the filament. This figure shows that gas from around the filament is being funneled onto the filament spine, and the gas along the spine is being funneled to the clumps which is promoting new star formation. This is in line with the observations of the accretion along the filament suggested in \citet{kirk} and simulations of cluster formation from hierarchical collapse (\citealt{vazquez}). This accretion is in line with the global hierachical collapse scenario presented in \citet{vaquez2019}. The accretion of filamentary gas by clumps located along the filament prevents the filament from forming new clumps because the formation of new clumps requires gas to collapse perpendicular to the filament. As seen by the streamlines in Figure \ref{fig:serpens_qual}, the gas is unable to collapse along the filament and instead moves parallel to the filament. 

We study the gas dynamics along the filament in more detail in Figure \ref{fig:serpens_dynamics}. This plot shows a snapshot of the dynamics while the filament is beginning to form new stars and stellar clumps at 0.2Myr. We first isolate the filament using \texttt{fil\_finder} (\citealt{fil_finder}) which gives us the coordinates of the filament spine and a mask enclosing the filamentary gas. We then rotate the reference frame by 25$^{\circ}$ to place the filament parallel to the new axes (x$^\prime$, y$^\prime$). Next, we calculate the velocity parallel to the filament spine (v$_{||}$) and the line of sight velocity (which, in the case of our simulations, is v$_z$) which are shown by the solid and dashed lines in Figure \ref{fig:serpens_dynamics} respectively. The first thing to notice from this figure are the spikes in velocity around the stellar clumps (stars are shown as red points). Clump 1 shows the strongest accretion as evidenced by the magnitude of the parallel velocity in its vicinity. South of the clump, the velocity is positive (implying upwards motion along y$^{\prime}$) but quickly transitions to negative (implying downwards motion along y$^{\prime}$) showing that clump 1 is accreting gas from both sides.

Moving to clump 2, we see a slightly different scenario. There is no transition from positive to negative parallel velocities around the clump but there is a peak in negative parallel velocity directly above the clump meaning that it is accreting from above in y$^{\prime}$. This lack of double-sided accretion is due to the presence of clump 1 which is more massive and as such has a wider pull of surrounding gas therefore diminishing the range of gas available for clump 2 to accrete. A similar case is seen for clump 3 where it is only able to accrete gas from directly above itself and not below in y$^{\prime}$. The parallel velocity directly below clump 3 is still negative implying that it is moving net towards clump 2.

Lastly, we note the existence of a peak in parallel velocity of the gas around y$^{\prime} = 0.4$pc that is not associated with a clump of stars. At $\approx 0.25$Myr, a small clump begins to form, but its growth is severely stunted. At this time, clumps one and two merge along the filament and begin to accrete the surrounding filamentary gas. Similarly, as clump 3 grows, its accretion becomes stronger. This leaves very little gas for the newly formed clump at y$^{\prime} = 0.4$pc to accrete before it merges with the remaining clumps. We discuss these mergers and the buildup of the cluster in more detail in Section \ref{sec:cluster_build_up}.

Accretion patterns are also seen along the line-of-sight as shown by the dashed line in Figure \ref{fig:serpens_dynamics}. In particular, we see that clumps one, two, and three are accreting along the line-of-sight. We also consider the effect of varying inclination angles. We find that, with inclination angles up to 10$^\circ$, we are able to clearly see the accretion signatures along the line of sight. In our simulation, accretion (both parallel to the filament and along the line-of-sight) is what leads to velocity gradients observed along the filament.

\subsection{Gas Accretion Onto Serpens South}
We also study gas accretion onto the filament itself from the surrounding gas rich environment at the same snapshot as Figure \ref{fig:serpens_dynamics}. We do this by considering two cylinders with radii of 0.1\,pc and 0.5\,pc around the filament in the x$^{\prime}$y$^{\prime}$ frame, and calculating the perpendicular velocity component (v$_{\perp}$) of the gas velocity within the annulus along the spine of the filament.

The accretion velocity onto the filament peaks at $\approx 0.85$km\,s$^{-1}$ around the location of clump 1 because of its high gravitational pull. Around clump 1, the accretion velocity does not change drastically until y$^{\prime} \approx 0.8$\,pc which is the location at which the filament becomes much more diffuse. It is also important to note that the accretion velocity of gas onto the filament is significantly lower than the parallel velocity along the filament feeding gas to the star forming clumps. As well, the mass accretion rate onto the filament is $\approx 7\times10^{-6}$M$_\odot$\,yr$^{-1}$ on average while that along the filament onto the clumps has several peaks $\gtrapprox 1\times10^{-4}$M$_\odot$\,yr$^{-1}$. This implies that the stellar clumps along the filament are being fed filamentary gas quicker than the filament is being fed gas from the surrounding environment. The Serpens South filament will run out of gas because of accretion onto the stellar clumps along its spine before it can replenish that gas from the surrounding environment.

We also compare the perpendicular velocity (v$_{\perp}$) to the velocity dispersion in the same component. The perpendicular velocity is a factor of 2.0-3.5 times higher than the velocity dispersion along the entirety of the filament indicating that turbulent motions are weak around the filament. It is important to note that this analysis would likely change with the inclusion of protostellar jets (e.g. \citealt{appel_2025}).

\subsection{Star Cluster Build Up}

\label{sec:cluster_build_up}
In this subsection, we discuss the build up of the final bound stellar system in the \texttt{fiducial} simulation. We consider the simulation finished when most of the bound gas has been removed from the system by stellar feedback or star formation. Gas is considered bound if its total energy is negative. By 1.6Myr, the remaining bound gas mass is $\approx$ 11\% of the initial gas mass meaning that the cluster is mostly gas free. We show the distribution of bound stars in the simulation at this time in Figure \ref{fig:final_snap}. Different clumps throughout the simulation are detected by applying the HDBSCAN (\citealt{hdbscan}) algorithm to the stars that are bound at any given time. We ensure that a group must contain at least 5 stars for HDBSCAN to consider it a cluster when discussing the early build up of the cluster along the Serpens South filament. These represent clumps that build up into the final star cluster. When using HDBSCAN to cluster the final system (as shown in Figure \ref{fig:final_snap}), we consider a system with 100 stars a cluster.

We begin by calculating the $Q$ parameter as defined in \citet{cart_Q} for the bound stars throughout the \texttt{fiducial} simulation. The $Q$ parameter is defined as $Q = \bar{m} / \bar{s}$ where $\bar{m}$ is the normalized mean length of the minimum spanning tree made for the bound stars in the simulation, and $\bar{s}$ is the mean separation between the bound stars divided by the radius of the cluster and is a measure of substructure in a distribution. The dividing value between substructured and smooth distributions is $Q = 0.8$ where values below 0.8 represent substructured distributions (fractal), and those above represent smooth ones. 

\begin{figure*}
    \centering
    \includegraphics[scale=0.25]{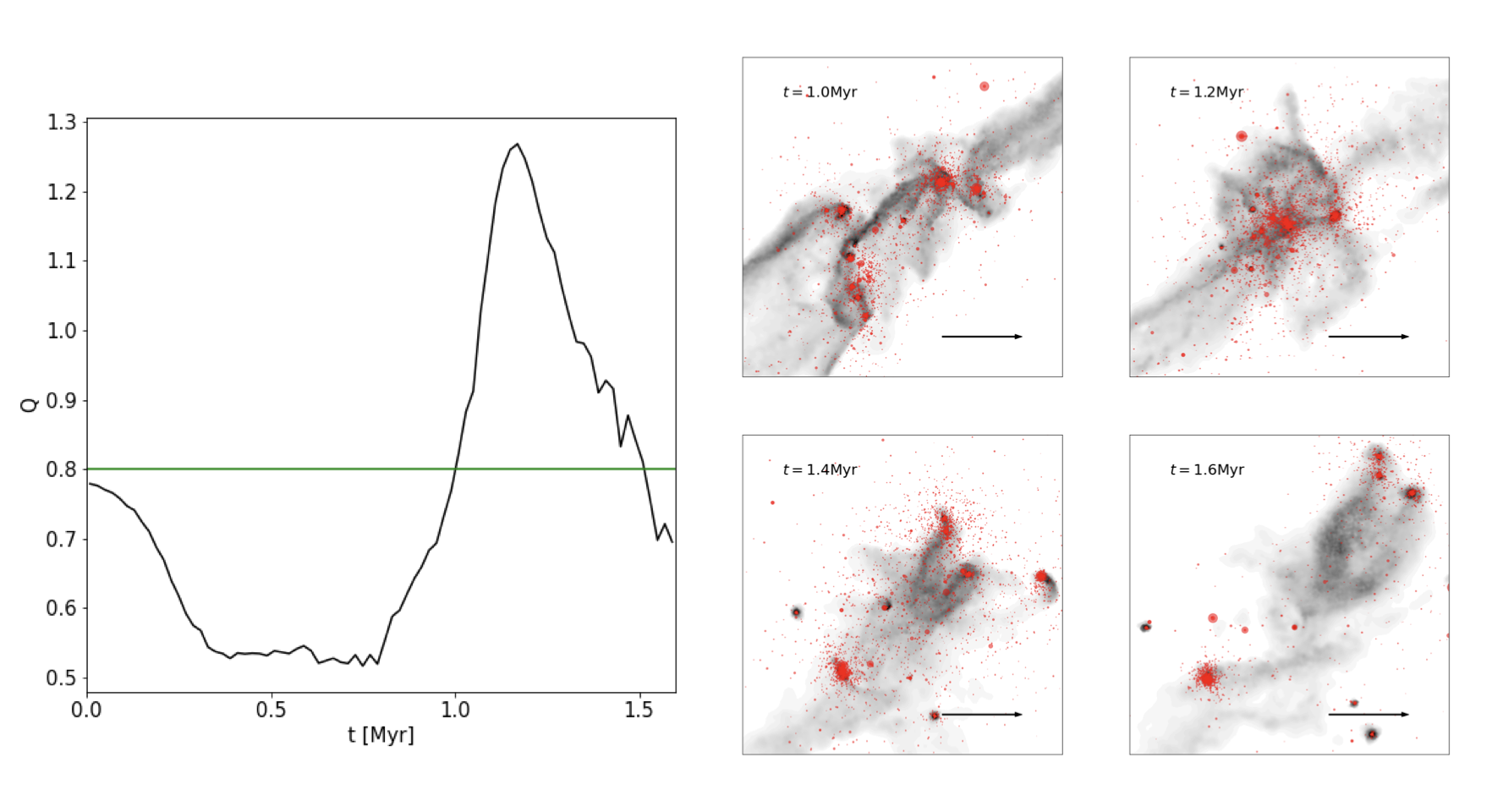}
    \caption{Left: The $Q$ parameter of the bound stellar component in the \texttt{fiducial} simulation. The green horizontal line indicates $Q = 0.8$ which is the threshold for fractal and smooth distributions. Right: Four snapshots showing the bound stars (red points) with their sizes scaled to the mass of the star the represent, and the gas distribution at $t = 1.0, 1.2, 1.4, 1.6$\,Myr. The arrow shows 1\,pc for scale.}
    \label{fig:q_param}
\end{figure*}

\begin{figure*}
    \centering
    \includegraphics[scale=0.32]{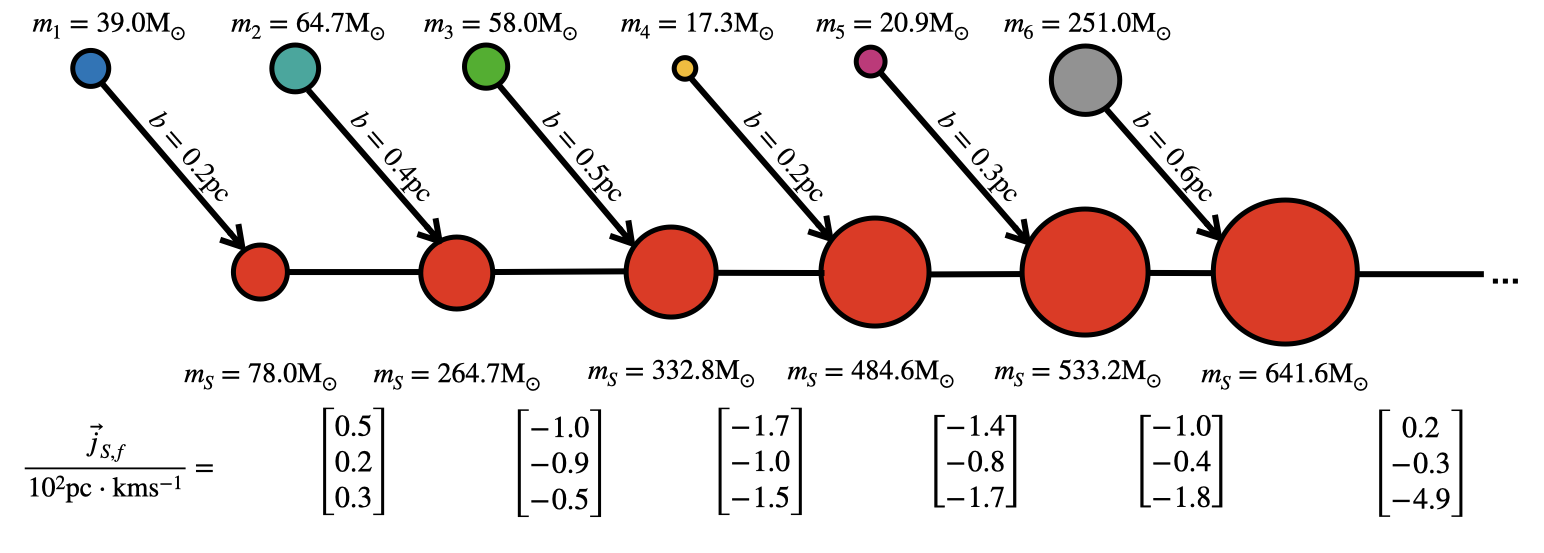}
    \caption{Schematic showing the mergers that make up the final Serpens South cluster in our \texttt{fiducial} simulation before it merges with W40. The bottom red circles show the main cluster and the top circles are different clusters that merged with it. The colours of the top circles are arbitrary and are used only to show that each merger is different. The size of all circles illustrate the mass of the cluster. The impact parameter ($b$) is shown for each merger along with the mass of the the main cluster at the start of the merger ($m_S$), the mass of the cluster it merges with ($m_{1,2,3,4,5,6}$) and the specific angular momentum vector of the main cluster after each merger is complete ($\vec{j}_{S,f}$).}
    \label{fig:merger_tree}
\end{figure*}

We show the evolution of the $Q$ parameter for the bound stars in the \texttt{fiducial} simulation in the left panel of Figure \ref{fig:q_param}. In the right panels, we show snapshots of the gas and bound stars at varying times during the final merger to illustrate the change in the $Q$ parameter. The bound stellar component begins around the threshold for smooth and substructured distributions but quickly begins to lower as star formation begins along Serpens South indicative of the clumpy star formation taking place along the filament. As the clumps along the filament begin to merge, the fractality decreases illustrated by a rising $Q$. We show the build-up of the final Serpens South cluster as a merger tree in Figure \ref{fig:merger_tree}. There are six mergers that make up the final Serpens South cluster each comprising of clusters of varying masses. The first cluster to form along Serpens South is the red circle in the bottom row of Figure \ref{fig:merger_tree} (this cluster grows from clump 1 in Figure \ref{fig:serpens_dynamics}). All of these mergers result in monolithic clusters with the exception of the merger of cluster six ($m_6$) with the main cluster ($m_S$). At $\approx 0.3$Myr after the merger of cluster six with the main cluster ($t$$\approx$1.1\,Myr), cluster six splits off. The early formation of a massive star could disrupt these mergers (\citealt{lewis_FB}).

All clusters that merge with the main cluster along Serpens South formed along Serpens South themselves as shown in Figure \ref{fig:serpens}, with the exception of cluster three ($m_3$) and cluster five ($m_5$). Cluster three forms at an overdensity in the initial condition located at ($\alpha$, $\delta$) $\approx$ (18$^{\mathrm{h}}$30.5$^{\mathrm{m}}$, -2$^{\circ}$.00$'$) and cluster five grows from an overdensity of YSOs located at ($\alpha$, $\delta$) $\approx$ (18$^{\mathrm{h}}$29.0$^{\mathrm{m}}$, -2$^{\circ}$.05$'$) before merging with the the main cluster along Serpens South. We discuss the dynamics of these mergers in more detail in Section \ref{sec:cluster_dynamics}. We find evidence of star formation triggered by the mergers of these clumps by looking at the star formation rate (SFR) throughout the simulation. For example, there is a peak in the SFR at $\approx 0.6$ free-fall times that corresponds with the monolithic collision of two dense cores along Serpens South.

At t$\approx$1.2Myr, the cluster that forms along Serpens South begins to merge with W40 (this causes the $Q$ parameter to reach a maximum indicating a smooth distribution as seen in Figure \ref{fig:q_param}). This final merger results in a cluster that is non-monolithic (illustrated by a decrease in the $Q$ parameter) and a fractal bound stellar system. The fractality post-merger arises because the merger between the Serpens South cluster and W40 is not head on, and the remaining bound gas mass in the region is not enough to drag the clusters back together. The stars originally belonging to the W40 cluster that remain bound by the end of the simulation are mostly part of the unclustered stars (gray points in Figure \ref{fig:final_snap}). The unclustered stars are also made up of stars that form elsewhere in the simulation (for example, along the Serpens South filament) but were dynamically ejected from their host clump. These are not runaway stars at this time because they remain bound in the simulation.

The final merger between W40 and the Serpens South cluster does not result in an increase in the SFR because the gas is mostly diffuse at this time. We note that this final merger between W40 and the Serpens South cluster is dependent on where we place W40 with respect to the Serpens south filament along the line of sight. If the two are disconnected by some distance, their potential merger would behave differently. This requires a change in modeling of the gas conditions and is out of the scope of this study. It will be addressed in future work.

\begin{figure*}
    \centering
    \includegraphics[width=1\linewidth]{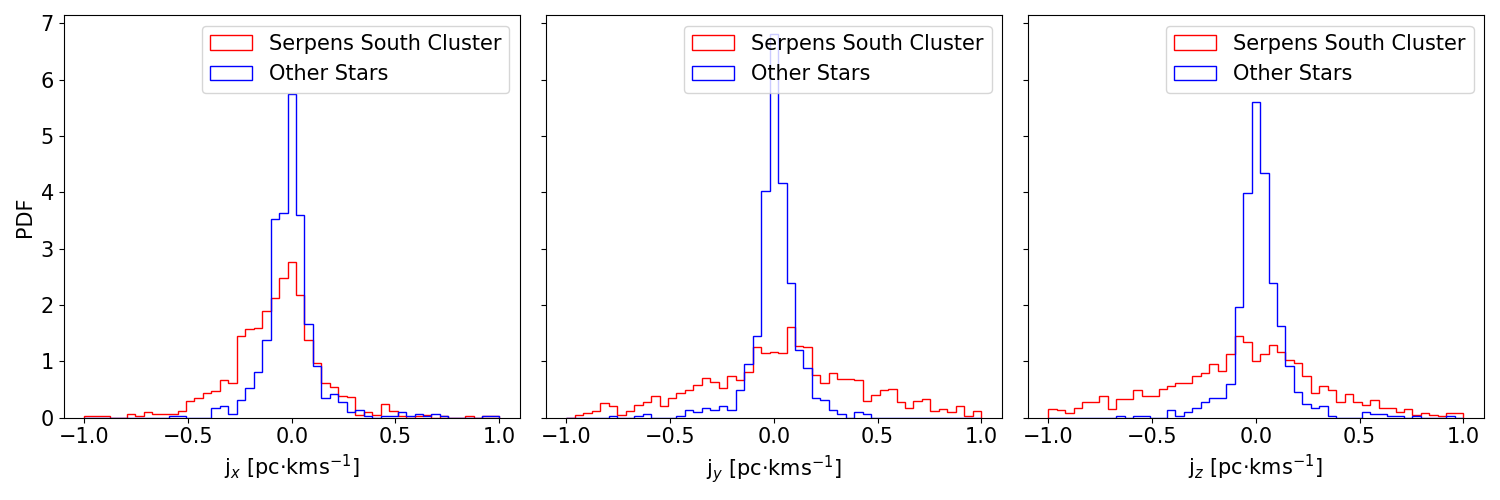}
    \caption{Distribution of specific angular momentum along the x (left), y (middle) and z (right) of the bound stars about the centre of mass of their respective subcluster at the end of the \texttt{fiducial} simulation. Red histograms show the distribution for bound stars belonging to the Serpens South cluster, and blue histograms show it for all other bound, clustered stars.}
    \label{fig:j_dist}
\end{figure*}

\begin{figure}
    \centering
    \includegraphics[scale=0.4]{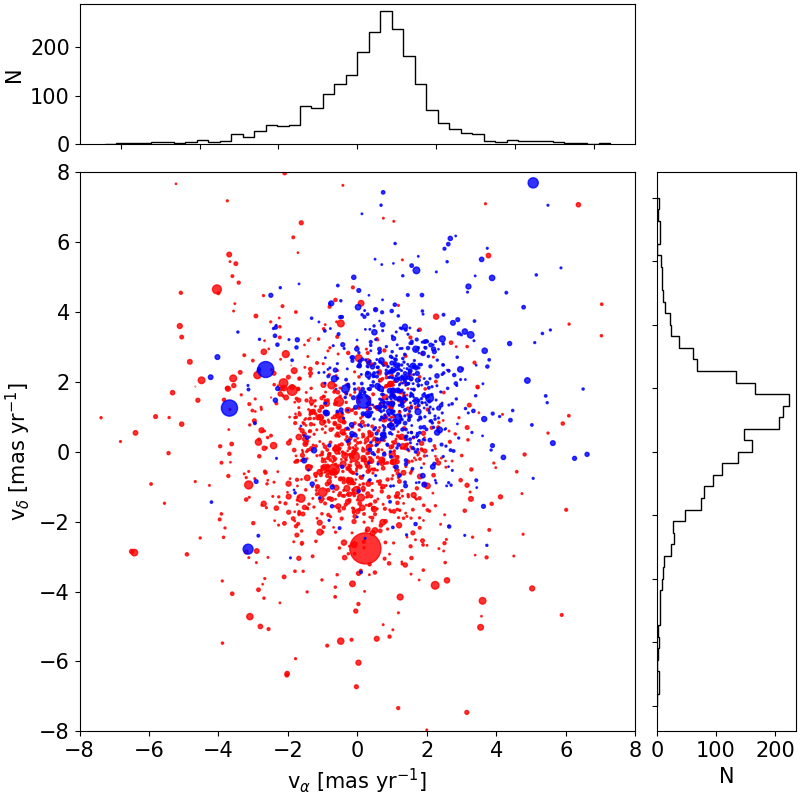}
    \caption{Distribution of bound stars in velocity space at the end of the \texttt{fiducial} simulation. The size of the points in the central scatter plot are proportional to the mass of the star. The red points show the Serpens South cluster stars, and the blue points show all other bound stars.}
    \label{fig:vel_space}
\end{figure}

\subsection{Cluster Dynamics}
\label{sec:cluster_dynamics}

The final bound stellar system in our simulation contains a cluster that assembles through multiple subcluster mergers along the Serpens South filament (red cluster in Figure \ref{fig:final_snap}, hereafter referred to as the Serpens South cluster), and a series of clusters that formed without substantial mergers (including W40). This provides us with a unique opportunity to study whether or not the dynamical signatures induced onto star clusters from mergers (see \citealt{Karam2024}), can still be seen after multiple mergers have taken place and to compare directly to clusters that formed in the same region merger-free. 

We begin with the dynamics inherited from star cluster assembly in the Serpens South filament. In Figure \ref{fig:merger_tree}, we show indicators of the dynamical evolution by showing the impact parameter ($b$) of each merger along with the specific angular momentum vector of the main cluster after each merger has finished. The impact parameter for each merger is calculated the same way as was done in \citet{Karam2024}: an impact parameter of 0pc is equivalent to a head-on collision.

All mergers that make up the final Serpens South cluster are not head-on. Because of this, the main cluster post-merger begins rotating. The first merger is counterclockwise in the $xy$ plane but the low mass of the clusters and the low impact parameter mean that the induced z-component specific angular momentum is only slightly positive. The remaining five mergers are all clockwise in the $xy$ plane resulting in the induction of increasingly negative z-component specific angular momentum in the final cluster. The behavior of the x and y-component specific angular momenta are more complicated. Beginning with the x-component, the mergers leading up to the final merger involving cluster six maintain negative specific angular momentum in the x direction. However, this changes when cluster six mergers counterclockwise in the $yz$ plane with the main cluster causing the system to have net positive specific angular momentum along the x direction. Lastly, the specific angular momentum along the y axis stays negative after the first merger because of the mergers with clusters two and three. However, it moves towards 0pc$\cdot$kms$^{-1}$ as the main cluster continues to merge with clusters four, five, and six in the opposite direction of its merger with clusters two and three. 

After the final non-monolithic merger between the Serpens South cluster and W40, we calculate the distribution of specific angular momentum of the Serpens South cluster. The specific angular momentum vector is $\vec{j}_{S,F}$ = ($-0.8$, $-0.5$, 0.4)$\times$10$^2$pc$\cdot$km\,s$^{-1}$ along the x, y, and z axes of the simulation. Comparing this vector with the specific angular momentum of the main cluster after merging with cluster six as shown in Figure \ref{fig:merger_tree} we see that the z component changes the most drastically. There are two reasons for this: the first is the removal of cluster six shortly after the merger and the second is because another cluster passes by the Serpens South cluster in the counterclockwise direction in the $xy$ plane which induces positive z-component specific angular momentum. 

We show the distribution of the specific angular momentum of bound stars by the end of the simulation in Figure \ref{fig:j_dist}. The multiple subcluster mergers that contribute to the formation of the Serpens South cluster have left their mark in the width of the specific angular momentum distribution along each axis. Conversely, clusters that have not gone through as many mergers have rotation distributions that are much thinner and centred on 0pc$\cdot$kms$^{-1}$. We find that using a different cluster centre and centre of mass velocity does not change the trends found in this simulation. If we use the average centre positions, or if we omit the stars in the outer-most region of the cluster (red points with x$>-0.5$pc in Figure \ref{fig:final_snap}), the distributions of the specific angular momenta of the Serpens South cluster stars are not as wide but remain wider than the specific angular momenta distributions for the remainder of the stars because of the hierarchical assembly of the Serpens South cluster.

Next, we show the distribution of all bound stars in the \texttt{fiducial} simulation in velocity space at the end of the simulation in Figure \ref{fig:vel_space}. Plotted are the velocities centered around the centre of mass velocity of the entire system and converted into units of mas/yr for easier comparison with observations. To make this conversion, we use the location of each star along the z-axis plus $d=436$\,pc. 

We see that the distribution of stars in velocity space is asymmetric. We also see this asymmetric distribution when looking at the radial velocities of the bound stars. It is important to note that though the distribution of stars in velocity space is asymmetric, it is not substructured (as can be seen by the black histograms). Violent relaxation (\citealt{tvr}) is the process responsible for removing kinematic substructure after cluster mergers. In cluster mergers, violent relaxation can cause relaxation of a clusters central component. However, the outer components of a cluster can remain unrelaxed (\citealt{arunima}) while remaining bound in gas-rich environments (\citealt{Karam2024}).

To test for deviations from normality, we perform the Shapiro Wilk test (\citealt{shapiro}) on the distributions of v$_\alpha$, v$_\delta$, and the radial velocities. We find that, for all distribution, the Shapiro Wilk test returns p-values significantly lower than 0.05 implying deviations from normality. We also consider whether or not the individual clumps that make up the final bound stellar system have non-Gaussian distributions in velocity space and find that each clump deviates from normality similar to the entire bound system. We calculate the velocity dispersion of each clump along v$_\alpha$, v$_\delta$, and v$_z$ (radial velocity). Of all clumps, the Serpens South cluster has the highest velocity dispersion along all three dimensions. This is because the Serpens South cluster was built up through mergers of smaller clumps and such mergers have been shown to result in increases in a clusters velocity dispersion (\citealt{Karam2024}, \citealt{karam_25}).

\begin{figure}
    \centering
    \includegraphics[scale=0.4]{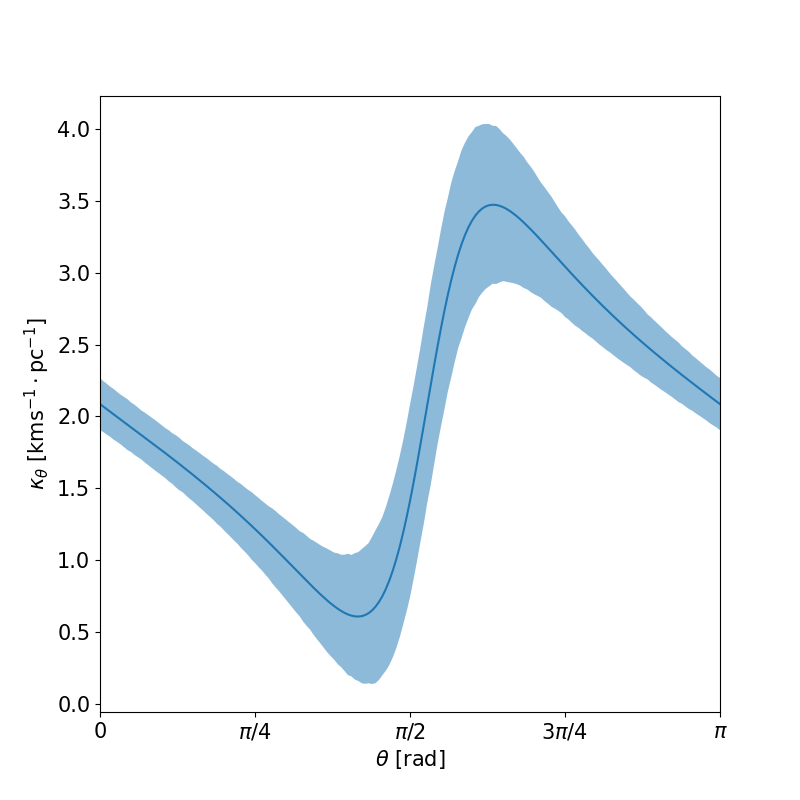}
    \caption{Expansion rate for the stars in the Serpens South cluster as a function of angle rotated about the Serpens South cluster centre at the end of the \texttt{fiducial} simulation. Shaded region shows 1$\sigma$. The errors in this figure are calculated by bootstrapping 10000 times.}
    \label{fig:expansion}
\end{figure}

We see from Figure \ref{fig:final_snap} that the entire bound stellar system is expanding by the end of the simulation. Along with this expansion, we test for expansion in each of the final clusters identified with HDBSCAN. We do this by fitting a line to the distribution of bound stars in each cluster in position-velocity space as shown in Figure \ref{fig:expansion} for the Serpens South cluster. The slope of the line ($\kappa_{\theta}$) gives the rate of the expansion along the axis of interest. To account for expansion throughout the entire cluster, we rotate the axis along which we calculate the expansion rate an angle $\theta$ above the x-axis from 0 to $\pi$ in increments of 1 radian and calculate the expansion rate along each axis for each cluster. For this analysis, we replace binaries with their centre of mass motions and velocities.

The Serpens South cluster is the only cluster that shows clear signs of expansion while expansion in the other clusters is dominated by uncertainty. This points to the importance of hierarchical build-up in driving early cluster expansion. The expansion rate as a function of $\theta$ for the Serpens South cluster is shown in Figure \ref{fig:expansion}. As expected for an anisotropically expanding cluster, the expansion rate varies depending on the angle of the axis considered. The angle $\theta$ with the strongest expansion rate ($\theta = 2$\,radians) is separated from the centre of mass velocity vector of the Serpens South cluster by an angle 0.9$\pi$/4. It is also separated from the initial Serpens South filament orientation by 0.9$\pi/4$. More simulations are needed to confirm a possible connection between merger axis and expansion axis.

\begin{figure}
    \centering
    \includegraphics[scale=0.4]{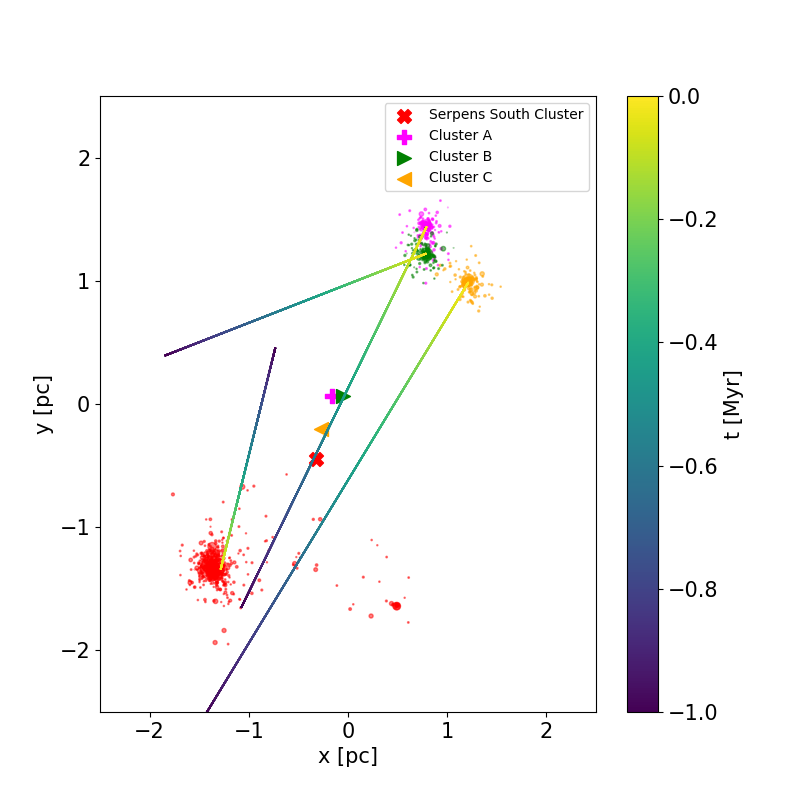}
    \caption{Trajectories of the clumps present at the end of the \texttt{fiducial} simulation traced back in time using the centre of mass and centre of mass velocities of each clump. Coloured points show the bound stars belonging to each clump (with the size proportional to the mass of the star it represents) and the lines show the trajectory in the $xy$ plane. The colour of the line shows the time since the simulations end at 1.6Myr. Markers show the location of the centre of mass of the same coloured cluster (See Figure \ref{fig:final_snap}) at the time of closest approach.}
    \label{fig:traceback}
\end{figure}

\subsection{Reconstructing Assembly History}

In this section, we trace back the motions of the individual stellar clumps in the final stellar system in the \texttt{fiducial} simulation to determine their point of closest approach. This process is shown in Figure \ref{fig:traceback}. Traceback has been used for dynamical age estimates of clusters in the Milky Way (e.g. \citealt{miret_roig2020}, \citealt{galli_2024}, \citealt{kerr_2025}) and by performing it on our simulation, we can compare the results to the known assembly history of the final stellar system to determine potential caveats to the traceback method as used for observations. Unlike observational studies, the contribution of the Milky Way potential is negligible in our simulation compared to the total potential of the system over the timescale considered. 

To calculate the time of closest approach using the traceback method, we invert the center of mass velocities of each clump along the plane of the sky ($xy$ plane in our simulation) and calculate the point at which the clumps were nearest one another. We find that all four clumps present at the end of the simulation were closest to one another $\approx 0.4$Myr before the end of the simulation (corresponding to t$=1.2$Myr). This corresponds well with the merger time for the merger of the Serpens South cluster with W40 that was determined using the $Q$ parameter (see Figure \ref{fig:q_param}). However, as shown by the crosses in Figure \ref{fig:traceback}, we see that the traceback method overestimates the distance of closest approach for the Serpens South cluster. This trend does not change when we redo the calculation with a different centre of mass and centre of mass velocity for each cluster. From Figure \ref{fig:traceback} we can see that the centre of mass velocity of the Serpens South cluster is slower on the plane of the sky ($xy$ plane) than the other clumps in the simulation. The reason for this comes from the geometry of the final cluster merger. At $\approx 0.2$Myr after the start of the merger, W40 and the Serpens South cluster reach their closest approach, but their centres do not merge resulting in the non-monolithic merger. At this point, the velocities along the x, y, and z axes are all at their peak and begin slowly decreasing throughout the remainder of the simulation. This illustrates that the complex kinematics of star cluster mergers could influence the assembly history of a cluster system that is derived using the traceback method.

We also note that the traceback method as employed here does not only imply that the clusters merged with one another in the past. Because it only tells us the time of closest approach, it could be that they emerged from the same molecular cloud and that their expansion is due to gas removal from stellar feedback. This is unlikely, however, because of the trajectory of the Serpens South cluster. In the case of expansion from gas removal, we would expect that the direction of motion of the Serpens South cluster should be directly opposite that of the other clumps in the simulation. This should especially be so if the closest approach was only 0.4Myr ago. 

Lastly, we see from Figure \ref{fig:traceback} that the traceback implies that the magenta and orange clumps were connected $\approx0.1$Myr ago. This is indeed the case as these used to be one cluster that eventually split. Cluster splitting has been seen in simulations of cluster assembly in molecular clouds (e.g. \citealt{dobbs_22}, \citealt{guszejnov2022}, \citealt{claude_2023}) and it is more likely to occur when gas mass is low because of the lower overall potential keeping the system bound (\citealt{karam}, \citealt{Karam2024}). In our simulation, this splitting occurs because the clusters are not bound to each other and, as the whole system expands, they grow further apart.

A more detailed analysis of the reliability of the traceback method will be addressed in Armstrong et al. (in prep).

\section{Comparison With Observations}
\label{sec:compare_obs}

\subsection{Gas}

The line of sight velocity gradients around Serpens South visible in Figure \ref{fig:serpens_dynamics} are similar to those found in \citet{kirk}. The three most defined line of sight velocity gradients in our \texttt{fiducial} simulation are leading to clump 1 from the south, connecting clump 2 to clump 3 and leading to clump 3 from the north. The first of these (leading to clump 1) was seen in \citet{kirk} and they determined a value for the flow rate of 1.4kms$^{-1}$pc$^{-1}$. In our simulation, the gradient is $\approx$1.8kms$^{-1}$pc$^{-1}$ very similar to the value found in \citet{kirk}. The accretion rate calculated in \citet{kirk} using line-of-sight velocities assumes that the Serpens South filament is inclined which leads to the visible gradient. We do not make this assumption in our initial condition but rather assume everything in the Serpens South filament is in the same plane within 0.1\,pc, yet we still see this velocity gradient appear along the line of sight. This occurs because, shortly after the beginning of our \texttt{fiducial} simulation, the Serpens South filament becomes inclined along the line of sight by a very small angle of $\approx7^\circ$ as determined using the stellar distribution along the filament.

\citet{kirk} also found infall velocities onto the filament of 0.25kms$^{-1}$ similar to our \texttt{fiducial} simulations values (peak of $\approx$0.85kms$^{-1}$ and decreases to $\approx$0.55kms$^{-1}$ away from the densest clump). As we did not include magnetic fields in our simulations, the similarity between our results and those from \citet{kirk} suggests that flows along Serpens South are gravitationally dominated near the filament spine. Further away from the spine, magnetic fields will become more important in shaping the kinematics of the gas (e.g. \citealt{suin_2025}).  

Lastly, we compare the mass accretion rates in our simulation to those presented in \citet{kirk} beginning with accretion along the filament long axis. We find an accretion rate of $\approx300$M$_\odot$Myr$^{-1}$ near the star forming clump 1 as defined in Figure \ref{fig:serpens_dynamics}. The trend of the accretion rate leading up to clump 1 is similar to that found in simulations (e.g. \citealt{smith_2016}, \citealt{wells_2025}) where the accretion rate peaks nearest the star forming clump and decreases with distance away from the clump. Away from the clump, the accretion drops to $\approx 50$M$_\odot$Myr$^{-1}$. The accretion in our simulation is higher than the \citet{kirk} accretion rate along the long axis calculated as $\approx30$M$_\odot$Myr$^{-1}$ but is in the range of accretion rates identified in other, more massive hub-filament systems (e.g. \citealt{hacar_2017}, \citealt{trevino_2019}, \citealt{he_g326_2023}). We note that our accretion rates act as an upper limit because we do not include protostellar feedback in our simulations. The discrepancy between the accretion rates of our simulation and the observations from \citet{kirk} likely comes from their assumed inclination angle of 45$^{\circ}$. If we instead convert their accretion rate value ($\approx30$M$_\odot$Myr$^{-1}$) to a value for a system inclined 7$^{\circ}$ using the method outlined in the \citet{kirk} paper, we obtain an accretion rate closer to our simulation of $\approx250$M$_\odot$Myr$^{-1}$.

We calculate a total accretion rate onto the southern filament of $\approx300$M$_\odot$Myr$^{-1}$ slightly higher than the \citet{kirk} value of $\approx 150$M$_\odot$Myr$^{-1}$. This discrepancy is likely a product of our use of multiple gas tracers when initializing the Aquila Rift.

\begin{figure}
    \centering
    \includegraphics[scale=0.4]{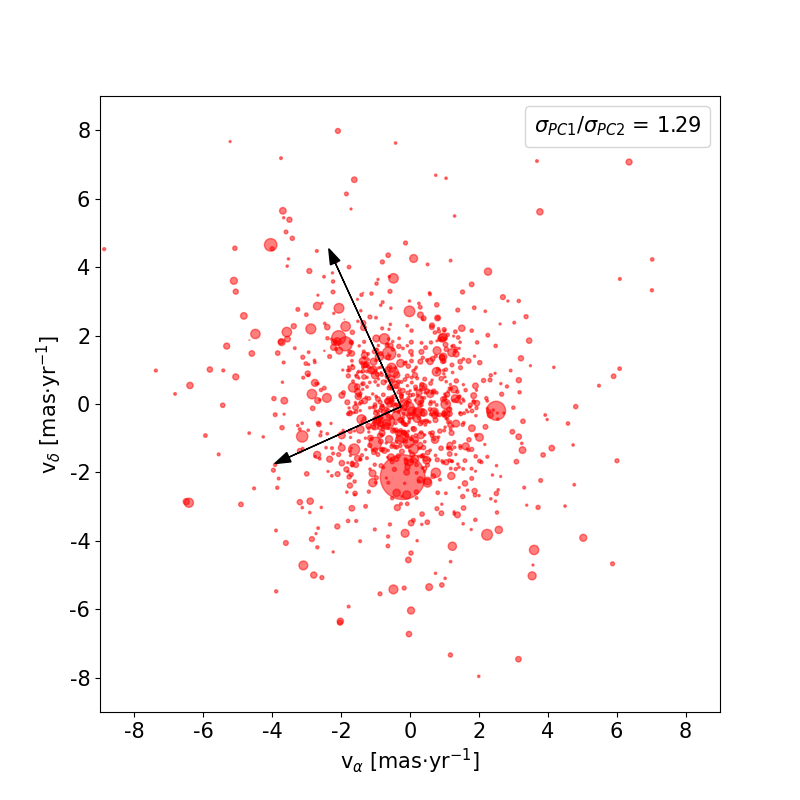}
    \caption{Velocity space distributions of bound stars at the end of the \texttt{fiducial} belonging to the Serpens South cluster. Arrows show the eigenvectors of the covariance matrix multiplied by their respective eigenvalues to illustrate the principal component axes. The ratio of the eigenvalues are shown in the top right.}
    \label{fig:sep_vel_space}
\end{figure}

\subsection{Star Clusters}
Observations from the \textit{Gaia} space telescope are providing us with detailed kinematics (proper motions along right ascension and declination, and radial velocities) of many young Milky Way clusters (e.g. \citealt{dias_2021}, \citealt{hunt_2023}) which are helping us understand the dynamical states (and evolution) of these clusters (e.g. \citealt{vazquez_2024}, \citealt{wright_2}). In this subsection, we compare the kinematic signatures from our simulations to these observations to better understand how observed clusters develop the kinematics they have.

We first discuss the velocity space distribution of stars in clusters. \citet{wright_2} used the Shapiro Wilk test to test for Gaussian velocity distributions in 18 stellar groups with ages ranging from 1Myr to 19.5Myr. They find that all clusters except for their oldest (Gamma Vel) show signs of deviation from normality similar to that found in our simulations. Anisotropies in velocity space have also been observed in the sample from \citet{kuhn_2019}. The authors find that the velocity space distributions of stars in their sample clusters and associations show signs of ellipticity in some cases as determined by fitting a bivariate normal distribution to the stars in velocity space and calculating the ratio of the long to the short axis. We perform the same velocity space analysis for the clusters in our simulation. We show an example in Figure \ref{fig:sep_vel_space}. Similar to \citet{kuhn_2019}, we find that the velocity dispersions can be heavily influenced by outliers in our simulations that are caused by binaries. To account for this, we replace all binaries present with their centre of mass velocities.

The ratios of the principal component eigenvalues for the clusters by the end of the \texttt{fiducial} simulation is similar for all four clusters with a value of $\approx$ 1.3. This value lies in the range found for the sample clusters in \citet{kuhn_2019} and, in particular, the Serpens South cluster (Figure \ref{fig:sep_vel_space}) matches closely with Trumpler 14 (Tr14) in the \citet{kuhn_2019} sample both in principal component ratios and orientation of the principal axes. Tr14 is part of the Carina Nebula and is aged at $\approx 1$Myr (\citealt{itrich_2024}) similar to our Serpens South cluster. A key difference, however, is that the values of the dispersion along each principal axis are higher for the Serpens South cluster in our simulation than for Tr14. The mass of Tr14 is estimated to be on the order of 10$^3$M$_\odot$ (\citealt{sana_2010}), double the final bound mass of the Serpens South cluster in our \texttt{fiducial} simulation. A virial analysis would suggest that the velocity dispersion of Tr14 should therefore be higher than our simulated Serpens South cluster. The multiple mergers experienced by the Serpens South cluster in our simulation have continually increased the velocity dispersion of the cluster while extending the total violent relaxation time required to relax from the merger. This may indicate that Tr14 did not form through subcluster mergers. Further towards this point, at t$\approx0.8$Myr, a star with mass 38M$_\odot$ appeared inside the Serpens South cluster, but the cluster was still nestled inside dense filamentary gas at this point. Conversely, Tr14 contains many O-type stars (\citealt{smith_2006}) whose feedback may be preventing it from experiencing future mergers or may have affected its ability to experience mergers in the past (\citealt{lewis_FB}). This occurs because of the change in gas morphology induced by feedback from massive stars which alters the motions of stellar clusters nestled within the gas and, in turn, the cluster merger histories. As shown in \citet{karam_25}, mergers lead to large increases in the velocity dispersion of a cluster compared to the clusters evolution post-merger with feedback considered.

The Carina Nebula as a whole is expanding (\citealt{goppl_2025}). As seen from Figure \ref{fig:final_snap}, our final bound stellar system is also expanding and contains a combination of clustered stars and more diffuse (yet still bound) field stars (shown in gray). \citet{goppl_2025} studied the expansion of each cluster in the Carina Nebula using the same method that we did in Section \ref{sec:cluster_dynamics}. The authors find that the expansion rate of stars in most of their clusters (including Tr14) peaks along one axis with only one cluster in their sample showing two peaks in the cluster expansion rate within error. This is similar to our results regarding the expansion of the Serpens South cluster in the \texttt{fiducial} simulation. A preference for directionality of high velocity unbound stars produced by subcluster mergers has been found in simulations (e.g. \citealt{CCC_JK}, \citealt{Polak2024}) and observations (e.g. \citealt{stoop_r136}).

\begin{figure}
    \centering
    \includegraphics[scale=0.4]{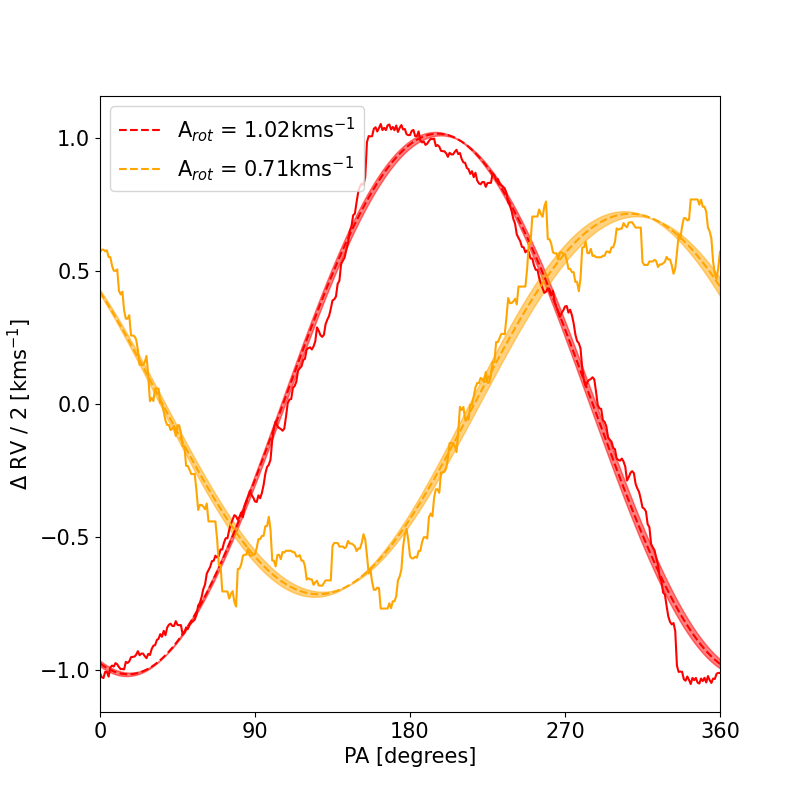}
    \caption{Half the difference in radial velocity as a function of the angle of the axis used to divide the stars in the clusters at the end of the \texttt{fiducial} simulation. Colours correspond to the clusters defined in Figure \ref{fig:final_snap} with the red line corresponding to the Serpens South cluster. Dashed lines show the best fit sine profile and the shaded regions show 1$\sigma$. The rotation amplitudes are shown in the top left.}
    \label{fig:obs_rotation}
\end{figure}

Anisotropic rotation has been observed in stellar associations (e.g. \citealt{wright_2016}) but observations from \citet{kuhn_2019} find little to no signatures of rotation in their sample of young clusters potentially due to high error. Recent work from \citet{jadhav_2024} has found signs of rotation in open clusters from the \citet{hunt_2023} sample. They do this by dividing a given cluster into halves separated by a line subtended by a position angle (PA) and calculating the mean velocity difference between these two halves. For a rotating cluster, a sinusoidal shape is expected when plotting the velocity difference against PA (e.g. \citealt{mackey_2013}). This has been seen for NGC 1846 (\citealt{mackey_2013}) and for ten clusters in the \citet{jadhav_2024} work. We perform this calculation for our clusters and show the results for the two clusters with the strongest rotation signatures from the \texttt{fiducial} simulation in Figure \ref{fig:obs_rotation}. For this analysis, we choose the cluster centre in a similar way as that done for the clusters in the \citet{hunt_2023} sample. We apply a kernel density estimation to the positions and velocities of the stars in each cluster and take the peak of the resulting distribution as the cluster centre of mass and centre of mass velocity.

In \citet{karam_25} we showed that mergers along with interaction with the surrounding gaseous environment can drastically change a clusters rotation which is what we are seeing here. The differing factor between the rotation in the Serpens South cluster (that built up hierarchically) and the other clusters (that did not build up hierarchically) is not the presence of rotation, but rather the distribution of that rotation among the stars that make up the cluster (as shown in Figure \ref{fig:j_dist}). Observational signatures of rotation in open clusters are still lacking (see \citealt{cantat_gaudin_2024} for a brief review) even with advances from \textit{Gaia}. Improved astrometry and an increase in radial velocity measurements in \textit{Gaia} data release 4 should make rotation measurements more plentiful.

Binary star clusters are pairs of open clusters that are connected with eachother in phase space and, in some cases, can be gravitationally bound to one another (e.g. \citealt{hu_2024}, \citealt{li_2025_BC}, \citealt{liu_2025}, \citealt{zhu_2025}). These systems can form from tidal capture of two clusters that formed in different clouds (\citealt{de_la_fuente_marcos_2009}), as well as through coeval formation in the same molecular cloud (e.g. \citealt{fujimoto_1997}, \citealt{bekki_2004}). Furthermore, some observed binary clusters have bridges of stars connecting the clumps (e.g. ASCC19 and ASCC20 as shown in \citealt{hu_2024}), but analysis of the dynamics of these stars is needed to determine their origin. Our simulations show that such systems can form as a result of the non-monolithic mergers of clusters that occur in mostly gas free environments. Eight of the sample thirteen binary clusters in \citet{li_2025_BC} were determined to be primordial binaries.

\section{Summary}
\label{sec:summary}
We perform star-by-star simulations of the Aquila star forming region by converting observations of the gas and stars in the region into simulation ready initial conditions. We use observations from \textit{Herschel}, \textit{Green Bank}, and \textit{Nobeyama} for the gas, and observations from \textit{Gaia}, the \textit{MYSTiX} survey, \textit{CFHT}, \textit{2MASS}, and the \textit{UKIDSS} survey for the stellar and pre-stellar component of the system. Star formation begins in unevenly spaced clumps along the Serpens South filament and these clumps eventually merge with one another to create a Serpens South cluster that merges with the already present W40 star cluster under our assumption that W40 and Serpens South exist at the same initial distance in the plane of the sky. The final merger is non-monolithic and results in a series of clumps rather than a single monolithic star cluster. The Serpens South cluster inherits unique dynamical signatures from its assembly history by the time of gas removal that are not present in the other clusters that form throughout the simulation. These include anisotropic expansion and a broadened distribution of angular momentum than the other clumps in the system. As well, the distribution of stars in velocity space is non-Gaussian and tilted similar to observations of open clusters in the Milky Way. The monolithicity of the final star cluster configuration is changes when we change the parameters of the initial conditions as discussed in the following Appendix.

In the future, we will simulate star cluster assembly in a wide range of observed clouds to further draw the connection between gas morphology and cluster assembly with a focus on how the gas morphology imprints itself on the dynamics of the forming cluster.

\begin{acknowledgments}
The authors thank Tomomi Shimoikura, Chi-Hong Lin, Rimpei Chiba, and Veronika Dornan for helpful discussions. The authors also thank the reviewer for constructive comments. JK is funded by the Japan Society for the Promotion of Science (JSPS) KAKENHI. This research was enabled by support provided by the Digital Research Alliance of Canada (https://www.alliancecan.ca/en).
\end{acknowledgments}

\software{astropy \citep{astropy:2013,astropy:2018,astropy:2022},  
          numpy \citep{numpy},
          matplotlib \citep{matplotlib},
          pynbody \citep{pynbody},
          Cloudy \citep{2013RMxAA..49..137F}, 
          scipy \citep{2020SciPy-NMeth}
          }


\appendix

\section{Varying Initial Conditions}
\label{sec:var_ics}
In this section, we discuss how our simulation initial conditions affect the results presented in Section \ref{sec:results}. The initial conditions are summarized in Table \ref{tab:sims}.

\subsection{Mass Resolution}
We start by comparing the results from our \texttt{fiducial} and \texttt{low\_res} simulations. The only difference between the initial conditions used for these simulations is the SPH particle mass (see Figure \ref{fig:ics}). In both simulations, the particle mass is enough to resolve Serpens South and the surrounding gas environment but the low particle mass in the \texttt{fiducial} simulation allows for better resolution of the density contrast throughout the region.

The general evolution of the simulation is the same for both initial conditions: star formation begins along the Serpens South filament, a cluster forms along the filament from the merger of smaller clumps, and this cluster eventually merges with W40. As such, the $Q$ parameter for the bound stars in each simulation behaves the same way up until $\approx 1$Myr. However, the merger between the Serpens South cluster and W40 ends differently in the two simulations. In the \texttt{low\_res} simulation, the merger between the Serpens South cluster and W40 is monolithic ($Q \approx 1.5$ at the time of the merger) converse to the non-monolithic merger between these two clusters at the end of the \texttt{fiducial} simulation. 

To investigate why this happens, we consider the mass of gas in each system. In the \texttt{low\_res} simulation initial condition, the gas mass is $\approx 12$\% ($\approx$1000M$_\odot$) higher than in the \texttt{fiducial} simulation. This is true because the cells in the \textit{Herschel} data that have masses (M$_\mathrm{cell}$) in between 0.0050M$_\odot$ and 0.0075M$_\odot$ will be assigned different total masses in each simulation because of the need to round M$_{\mathrm{cell}}$/$m_{\mathrm{SPH}}$ to an integer. In the \texttt{low\_res} case, they will be assigned a mass two times greater than in the \texttt{fiducial} case leading to an overall increase in the total mass of the system. This added mass leads to the formation of a more massive bound stellar system by the time of the merger between the Serpens South cluster and W40 compared to that in the \texttt{fiducial} simulation. An increase in total mass such as this leads to a deeper gravitational potential well that acts to keep the two clusters together post-merger. If we performed simulations with even higher SPH particle mass (lower mass resolution), we risk not resolving the more diffuse medium around Serpens South which would also change the overall behavior of the simulation.


The distribution of stars in velocity space at the end of the \texttt{low\_res} simulation is similar to that of the \texttt{fiducial} simulation: non-Gaussian according to the Shapiro Wilk test. However, because the final cluster in the \texttt{low\_res} simulation is considered monolithic, the distribution of the stars in velocity space is more symmetric than it was in the \texttt{fiducial} model. 

The final monolithic cluster in the \texttt{low\_res} simulation has a non-bivariate distribution in the velocity space distribution of the bound stars, similar to the \texttt{fiducial} simulation, but with much higher overall velocity dispersion. This is a result of the added velocity dispersion induced by the final merger between the Serpens South cluster and W40 in this simulation and the higher mass of the final cluster. As well, the final cluster present at the end of the \texttt{low\_res} simulation is expanding anisotropically.

\subsection{Observational Datasets}
In this subsection, we compare our \texttt{fiducial} simulation to our \texttt{no\_NH3} simulation to discern the role played by the inclusion of observations of dense gas tracers in our simulation initial conditions. The first major difference is in the time at which gas collapses to form stars. Because we are only using $^{13}$CO dynamics (which traces more diffuse gas) in our \texttt{no\_NH3} model, the gas overall has higher velocity dispersion and therefore takes longer to collapse and begin star formation. However, once star formation begins, the amount of bound stellar mass formed throughout the \texttt{no\_NH3} simulation is similar to that of the \texttt{fiducial} simulation.

The $Q$ parameter in the \texttt{no\_NH3} simulation evolves similarly to that in the \texttt{fiducial} simulation but does not go below the threshold value of $Q=0.8$ by the end of the simulation. Converse to the \texttt{fiducial} simulation, the buildup of the Serpens South cluster in the \texttt{no\_NH3} simulation involves many more non-monolithic mergers of clumps along the Serpens South filament. Because of this, the Serpens South cluster is never able to fully coalesce into a monolithic cluster before merging with W40. This leads to the destruction of the small clumps that make up the Serpens South cluster shortly after its merger with W40 and the resulting bound stellar mass is significantly lower ($\approx500$M$_\odot$) than in the \texttt{fiducial} of \texttt{low\_res} simulations. As a result, HDBSCAN does not identify any clusters in the final bound stellar distribution of the \texttt{no\_NH3} simulation.

The reason for the non-monolithic mergers building up the Serpens South cluster is not necessarily tied to the lack of high-density gas data used in initializing the simulation. As the clumps form stars along the filament, binary and multiple systems start to appear and their internal dynamics can dislodge the clump from the host filament meaning that the merger of these clumps is off-axis. We see this happen in the \texttt{low\_res} simulation as well but because the mass is higher in the \texttt{low\_res} simulation overall, the clumps stick together even though the mergers are non-monolithic. This implies that the presence of binaries, especially early on in the cluster assembly process, can be quite important in determining a clusters final configuration. Future work will involve simulation suites of cluster assembly with the inclusion of primordial binaries to fully assess their importance.


\bibliography{sample701}{}
\bibliographystyle{aasjournalv7}



\end{document}